\documentclass[structabstract]{aa}  
\usepackage{graphicx}
\usepackage{txfonts}
\usepackage{natbib}
\bibpunct{(}{)}{;}{a}{}{,}

\defcitealias{cg1head}{Paper~I}
\defcitealias{bessellbrett88}{BB88}
\newcommand{\twco}{{\hbox {\ensuremath{\mathrm{^{12}CO}} }}}
\newcommand{\ceo}{{\hbox {\ensuremath{\mathrm{C^{18}O}} }}}

\newcommand{\thco}{{\hbox {\ensuremath{\mathrm{^{13}CO}} }}}

\newcommand{\Msun}{\ensuremath{\mathrm{M}_\odot}}

\newcommand{\Lsun}{\ensuremath{\mathrm{L}_\odot}}

\newcommand{\hour}{\ensuremath{^\mathrm{h}}}
\newcommand{\minute}{\ensuremath{^\mathrm{m}}}
\newcommand{\second}{\ensuremath{^\mathrm{s}}}

\newcommand{\mum}{\mu {\rm m}}
\def\Ks{\hbox{$K$s}}

\def\Js{\hbox{$J$s}}
\def\J{\hbox{$J$}}
\def\H{\hbox{$H$}}
\newcommand{\umag}{{$^{m}$}}

\def\jshks{\hbox{$J$s$\!H\!K$s}} 
\def\jhks{\hbox{$J\!H\!K$s}}              % JHKs system
\newcommand{\ceoSE }{{\hbox {\ensuremath{\mathrm{C^{18}O}}\_SE}}}

\newcommand{\Htwo}{{\hbox {\ensuremath{\mathrm{H_2}}}}}

\def\fm{\hbox{$.\!\!^{\rm m}$}}
\begin{document}
   \title{Star formation, structure, and formation mechanism of cometary globules: NIR observations of CG~1 and CG~2
   \thanks{Based partly on observations done at the European Southern
       Observatory, La Silla, Chile (ESO programme 078.C-0490).}\fnmsep
       \thanks{Appendices A and B are only available in electronic form at http://www.aanda.org}}

   \author{M. M. M\"akel\"a
          \inst{1}
          \and
          L. K. Haikala\inst{2,1}
          }

   \institute{Department of Physics, Division of Geophysics and Astronomy,
             P.O. Box 64, 00014 University of Helsinki, Finland\\
              \email{minja.makela@helsinki.fi}
   \and
              Finnish Centre for Astronomy with ESO (FINCA),
              University of Turku, V\"ais\"al\"antie 20, 21500 Piikki\"o, Finland 
             }

   %\date{Received 17 July, 2012/Accepted 22 October 2012}
    \date{}

% \abstract{}{}{}{}{} 
% 5 {} token are mandatory
 
  \abstract
   {Cometary globule (CG) 1 and CG~2 are ``classic'' cometary globules in the Gum Nebula. They have compact heads and long dusty tails that point away from the centre of the Gum Nebula.}
   {We study the structure of CG~1 and CG~2 and the star formation in them to find clues to the CG formation mechanism. The two possible CG formation mechanisms, radiation-driven implosion (RDI) and a supernova blast wave, produce a characteristic mass distribution where the major part of the mass is situated in either the head (RDI) or the tail (supernova blast).}
   {CG~1 and CG~2 were imaged in the near infrared (NIR) \jshks\ bands. NIR photometry was used to locate NIR excess objects and to create visual extinction maps of the CGs. The $A_{V}$ maps allow us to analyse the large-scale structure of CG~1 and CG~2. Archival images from the WISE and Spitzer satellites and HIRES-processed IRAS images were used to study the globule's small-scale structure. Fits were made to the spectral energy distribution plots of the NIR-excess stars to estimate their age and mass.}
   {In addition to the previously known CG~1 IRS~1 we discovered three new NIR-excess objects in IR imaging, two in CG~1 and one in CG~2. CG~2 IRS~1 is the first detection of star formation in CG~2. The objects are young low-mass stars. CG~1 IRS~1 is probably a class I protostar in the head of CG~1. CG~1 IRS~1 drives a bipolar outflow, which is very weak in CO, but the cavity walls are seen in reflected light in our NIR and in the Spitzer 3.6 and 4.5 $\mum$ images. Strong emission from excited polycyclic aromatic hydrocarbon particles and very small grains were detected in the CG~1 tail. The total mass of CG~1 in the observed area is 41.9 \Msun\ of which 16.8 \Msun\ lies in the head. For CG~2 these values are 31.0 \Msun\ total and 19.1 \Msun in the head. 
The observed mass distribution does not offer a firm conclusion for the formation mechanism of the two CGs: CG~1 is in too evolved a state, and in CG~2 part of the globule tail was outside the observed area.}
   {Even though the masses of the two CGs are similar, star formation has been more efficient in CG~1. By now, altogether six young low-mass stars have been detected in CG~1 and only one in CG~2. A possible new outflow was discovered to be emanating from CG~1 IRS~1.
   }


   \titlerunning{NIR observations of CG~1 and CG~2}
   \authorrunning{M. M. M\"akel\"a \& L. K. Haikala}

\maketitle

\section{Introduction} \label{sect:introduction}

Star formation takes place in dense molecular clouds when gravity starts dominating over the gas pressure. In triggered star formation this first ``kick'' to create a density fluctuation that will lead to the cloud collapse can be provided by mechanisms such as stellar winds or supernova explosions \citep[e.g.][]{elmegreen98}. The collapse continues until the conditions in the collapsing cloud are suitable for star formation and the gas pressure starts to resist further collapse. Small isolated molecular clouds and globules, such as the ones described by \citet{bokreilly47}, are ideal locations to study the star formation process in detail. The advance of large-format NIR arrays and especially of space-borne telescopes like the Spitzer Space Telescope has made it possible to detect deeply embedded low-mass young stellar objects in seemingly starless dark clouds \citep [see e.g][]{youngetal04}. These embedded objects are not visible at shorter wavelengths because of the high extinction. If the formation of these stars was triggered by some outside event, the triggering mechanism may have left a telltale sign into the large-scale cloud structure. Determining the triggering mechanisms leads to a better understanding of star formation processes and the characteristics of newly formed stars. 

Cometary globules (CGs) are a special class of globules. CGs have dusty, compact ``heads'' and elongated, faintly luminous ``tails''. They were first discovered in the Gum Nebula \citep{hawardenetal76, sandqvist76}. CGs are typically found near OB associations in HII regions \citep[e.g. Rosette;][]{whiteetal97}. Some of the ``classical'' CGs in the Gum Nebula are sites of isolated low-mass star formation \citep[e.g.][]{reipurth83}.

Two formation mechanisms for CGs have been suggested. 
\citet{brandetal83} argued that CGs form when a supernova (SN) explosion shocks an originally spherical cloud. The shock compresses the cloud to form the head, and the blast wave drives the material mechanically away from the SN to form the tail. \citet{reipurth83} suggested radiation driven implosion (RDI) where the UV radiation from massive O stars photoionizes the cloud and shock fronts compress the cloud. The less dense cloud medium is separated from the cloud by radiation and ionization shocks, and the tail is formed out of the eroded cloud medium.
Numerical simulations have been done for both mechanisms and they show that both RDI and SN can create tails behind the globule core.
According to the RDI model, a major part of the CG mass is in the ``head'' \citep[e.g.][]{kesseldeynetetal03, leflochlazareff94, bertoldimckee90}, whereas in the SN model a large part of the mass can lie in the tail \citep[e.g.][]{heathcotebrand83, bedogniwoodward90}. This offers the possibility of determining the mechanism that triggers the formation of CGs by studying the mass distribution in them.

We used NIR imaging to study the structure of and star formation in two archetype cometary globules, \object{CG~1} and \object{CG~2}, in the Gum Nebula. These globules are a part of the system of cometary globules in the Gum Nebula \citep{reynolds76}. This nebula has been described as a supernova remnant and/or an HII region.
A SN explosion $\sim$1 Myr ago has been suggested as the possible origin of the Gum Nebula \citep[e.g.][]{reynolds76b, brandetal83}. As an HII region, the Gum Nebula is mainly ionized by two O stars, the multiple system $\gamma^{2}$ Velorum and $\zeta$ Puppis. The tails of the CGs point away from the Gum Nebula centre, suggesting a common triggering mechanism \citep{hawardenetal76, zealeyetal83}. CG~1 is one of the largest of the Gum Nebula CGs. Its tail is 25\arcmin\ long and the characteristic size of the globule head is 2\arcmin\ \citep{brandetal83}. CG~2 is a smaller CG situated $\sim$52\arcmin\ northwest (NW) of CG~1. Even though the tail length is listed as 26\arcmin\ by \citet{zealeyetal83} and 18\arcmin\ by \citet{sridharan92}, the effective tail length is $\sim$11\arcmin\ and the rest is only marginally detected in the optical. The distance estimates to CG~1 range from 300 to 500 pc, and this paper adopts the value 300 pc of \citet{franco90} used in \citet{cg1head} \citepalias{cg1head}. 

The observations of the CG~1 head have been discussed in \citetalias{cg1head} where the detection of a young stellar object embedded in the dense globule head was reported. This paper expands the observations to cover the CG~1 tail and the nearby CG~2.
Observations and data reduction are described in Sect. \ref{sect:observations}, and the results are presented in Sect. \ref{sect:results}. Discussion is in Sect. \ref{sect:discussion} and a summary in Sect. \ref{sect:conclusions}.

\section{Observations and data reduction} \label{sect:observations}

\subsection{NIR Imaging} \label{sect:imagingdata}

CG~1 was imaged in \J, \H, and \Ks\ with the Simultaneous InfraRed Imager for Unbiased Survey \citep [SIRIUS,][]{nagayamaetal03} on the InfraRed Survey Facility (IRSF) 1.4m telescope at SAAO in Jan. 2007. 
The SIRIUS field of view is 7\farcm7$\times$ 7\farcm7 and the pixel scale 0\farcs453. The observations were carried out in the on-off mode to preserve the source surface brightness.  The off-position was at 7\hour18\minute\  $-44^{\mathrm o}$05\arcmin, (J2000). Each on and off observation consisted of ten individual jittered images. Depending on the stability of the weather 10 s or 15 s integration time was used for the individual images. The average seeing was 1\farcs3--1\farcs4. Four overlapping fields were observed to cover the area where \twco was observed by \citet{harjuetal90}. Going from east to west, the fields will be referred to as I(RSF)1, I2, I3, and I4. The observing sequence was off - I1 - I2 - off - I3 - I4 - off, and thus the total integration time ($\sim$1500 s) for each field is the same. Sky flats were observed in the evening.
CG~2 was imaged in the same way as CG~1, but only two overlapping fields were observed. The total integration times of the fields are $\sim$2200 s and $\sim$1200 s.

CG~1 was imaged in \Js, \H, and \Ks\ with SOFI (the Son of Isaac) near-infrared instrument on the New Technology Telescope (NTT) at the La Silla Observatory, Chile. The SOFI field of view is 4\farcm92 $\times$ 4\farcm92 and the pixel size is 0\farcs288. The imaging covered the area  where \citet{harjuetal90} detected \thco emission. Three overlapping SOFI fields were observed. The fields will be referred to as N(TT)1, N2 and N3 (from east to west). The observations were carried out in Feb. 2007 except for the field N3 in \Ks, which was observed in Jan. 2010. 
SOFI imaging was done in the standard jitter mode with jitter box widths of 20\arcsec--30\arcsec. Fields N1 and N2 were imaged in observation blocks of 50 individual images of one minute integration each. Three and two observing blocks in all colours were obtained for N1 and N2, respectively. One observing block of 25 images in the three colours was obtained in field N3 and in the same off-field which was used in the IRSF observations. Standard stars from \citet{perssoncatalog} were observed before and after each observation block. CG~2 was imaged with SOFI in \Js\ and \Ks\ bands in Feb. 2007 and in \H\ band in Jan. 2010. The observations were done in the standard jitter mode with a jitter box width of 30\arcsec. Images were obtained in observation blocks of 25 individual frames with one minute integration each. \Js\ and \H\ bands have one and \Ks\ band has two observation blocks. The average seeing during the SOFI observations was 0\farcs8.

\subsection{Data reduction} \label{sect:datareduction}

The data for SIRIUS and SOFI was reduced as described in \citetalias{cg1head}. 
For the sake of completeness, the data reduction steps for the single fields are included in Appendix \ref{app:datareduction}. Data from the separate frame catalogues were combined into one. Any duplicate stars from the overlap regions were averaged into a single entry.

The final catalogues for SIRIUS fields contain 106 objects (I1), 161 (I2), 218 (I3) and 172 (I4). For the SKY field, the final catalogue has 313 objects. The CG~2 catalogues have 337 (I1) and 186 objects (I2). The final combined SIRIUS catalogue for CG~1 and CG~2 have a total of 576 and 466 stars, respectively.

The number of objects in SOFI catalogues for CG~1 are 261 in N1, 296 in N2 and 142 in N3. The combined photometry catalogue of CG~1 has 674 stars, the single CG~2 frame has 201 and the off-field 230 stars. The stellar limiting magnitudes corresponding to a 0\fm1 error in the SOFI catalogues for N1 and N2 are 20\fm5 in \H\ and \Ks\ and 21\fm5 in \Js. For N3, OFF, and CG~2 the limiting magnitudes are $\sim$0.5-1.0\umag\ brighter.

\section{Results} \label{sect:results}

\subsection{NIR Imaging} \label{sect:imagingresults}

\subsubsection{CG~1} \label{sect:results_cg1im}

A false-colour image of the SIRIUS data combined from on-off-mode observations is shown in the online Fig. \ref{fig:cg1_3colorsirius}. The \J, \H, and \Ks\ bands are coded in blue, green, and red, respectively. The brightest object in the image is \object{NX Pup} at 7\hour19\minute28.3\second\ $-44^{\mathrm o}$35\arcmin 11\arcsec. 
A bright reflection nebula (RN) is seen below NX Puppis in the figure and is most probably illuminated by it. A thin, bright filament leads southwest (SW) from the RN and a fainter filament extends NW from the nebula. The reddish, semi-nebulous bright patch 70\arcsec\ west of NX Pup is due to reflected light from a newly born star (young stellar object; YSO) embedded in the dense globule head. Extended surface brightness is observed NW of the YSO. A bright arc reaches northeast (NE) from the YSO. See \citetalias{cg1head} for discussion on the detailed structure of the head. About 6\arcmin\ west of the YSO lies another region of extended surface brightness. The on-off observation mode preserves the true surface brightness and therefore the extended structure seen in the online Fig. \ref{fig:cg1_3colorsirius} is real. 

\begin{figure*}[ht]
\centering
\includegraphics [width=17cm] {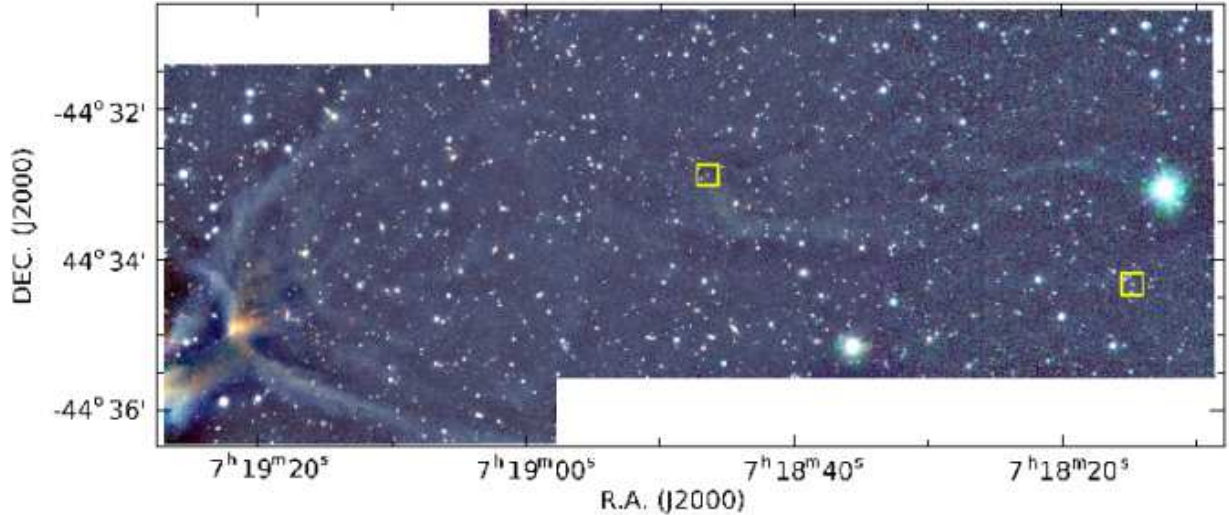}
   \caption{Colour-coded SOFI image of CG~1. The \J, \H, and \Ks\ bands are coded in blue, green, and red. The objects CG1 IRS~2 and 3 (from east to west) are indicated with a box.}
\label{fig:cg1_3colorsofi}
\end{figure*}

A false-colour image of CG~1 combined from SOFI images is shown in Fig. \ref{fig:cg1_3colorsofi}. The objects CG~1 IRS~2 and 3 (see  Sect. \ref{sect:starformation}) are indicated. NX Pup is too bright to be observed with SOFI and was therefore left outside the image. The edge of the RN below NX Pup is seen in the lower left-hand corner of the figure. The filamentary structure and the YSO are seen more clearly than in Fig. \ref{fig:cg1_3colorsirius}. The jittering observing mode uses the observed ON frames to estimate the sky brightness. This smears out surface brightness structures larger than the jitter box width and produces filaments that trace gradients in the original background. Point sources and galaxies are unaffected. Therefore the extended surface brightness NW of the YSO seen in the online Fig. \ref{fig:cg1_3colorsirius} has largely disappeared, but the northern filament is seen more clearly because of strong local surface brightness variation. The negative brightness areas right next to the filaments are not real but due to the jittering. The overall morphology of CG~1 in the SOFI data can be best seen in the \Js\ and \H\ bands (online Fig. \ref{fig:cg1_sofiall}, upper panels). In \Ks\ (online Fig. \ref{fig:cg1_sofiall}, lower panel), only the leading edges, a short filament behind them, and the brightest tail filament are seen. Most of the extended surface brightness NW from the YSO seen in the SIRIUS image has been transformed into a faint filamentary structure. In the western part of the tail, another fainter, slightly arced filament is seen.

\subsubsection{CG~2}
\label{sect:results_cg2im}

\begin{figure}
\resizebox{\hsize}{!}{
\includegraphics {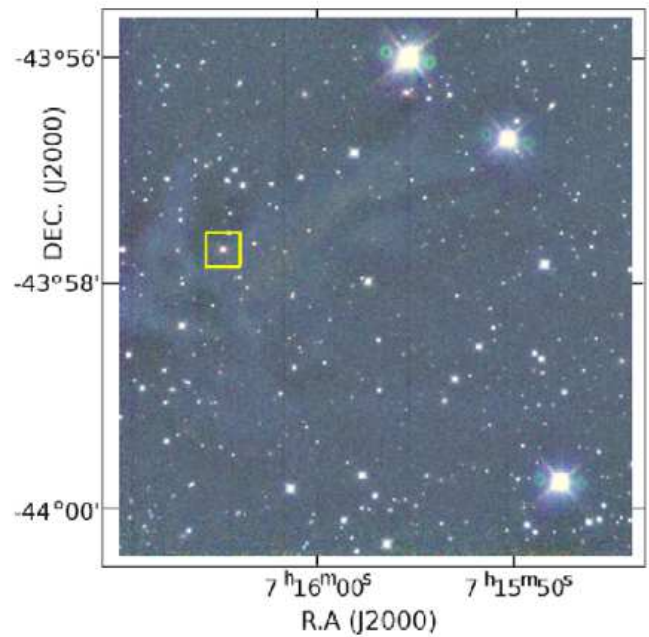}
}
   \caption{Colour-coded SOFI image of CG~2. The \Js, \H, and \Ks\ bands are coded in blue, green, and red. The object CG~2 IRS~1 is indicated with a box.}
\label{fig:cg2_3colorsofi}
\end{figure}

A false-colour image of CG~2 obtained with SIRIUS in the on-off observing mode is shown in the online Fig. \ref{fig:cg2_3colorsirius}. The \J, \H, and \Ks\ bands are coded in blue, green, and red, respectively (the individual SIRIUS \J, \H, and \Ks\ images are shown in the online Fig. \ref{fig:cg2sirius_all}). The object CG~2 IRS~1 (see  Sect. \ref{sect:starformation}) is indicated with a box in the image. In \J\ (online Fig. \ref{fig:cg2sirius_all}, top left panel), extended surface brightness is observed in the globule head. There is a ``nose'' of surface brightness just east of a more extended region of enhanced surface brightness where visibly reddened stars are located. No details of the tail can be distinguished in the SIRIUS images. Only the NE part of the online Fig. \ref{fig:cg2_3colorsirius} is free of extended surface emission, and the sky to south and to east of the globule head is covered by faint surface brightness.

The SOFI image of CG~2 is shown in Fig. \ref{fig:cg2_3colorsofi} in false colour (the individual \Js, \H, and \Ks\ images are in the online Fig. \ref{fig:cg2_sofiall}). The object CG~2 IRS~1 is marked. The ``nose'' appears slightly bluish. Behind the nose, a filamentary structure produced by the jittering observing mode extends to the NW. Reddened stars can be seen in this extended region. A faint filament runs along the SW edge of the globule. The CG~2 tail is not seen because it extends far outside the $\sim$5\arcmin\ field of view.

\subsection{Photometry} \label{sect:photometry}

The SOFI limiting magnitude in each colour is $\sim$2\umag\ brighter than in the SIRIUS data. The SOFI data is discussed in detail in the following. Galaxies should have been removed from the catalogue by the selection rules listed in the online Appendix \ref{app:datareduction}, but it is likely that some still remain. The catalogues containing the SOFI photometry for CG~1, CG~2, and the OFF region are available at the CDS.

\subsubsection{CG~1}
\label{sect:results_cg1ph}

\begin{figure*}
\centering
\includegraphics [width=17cm]{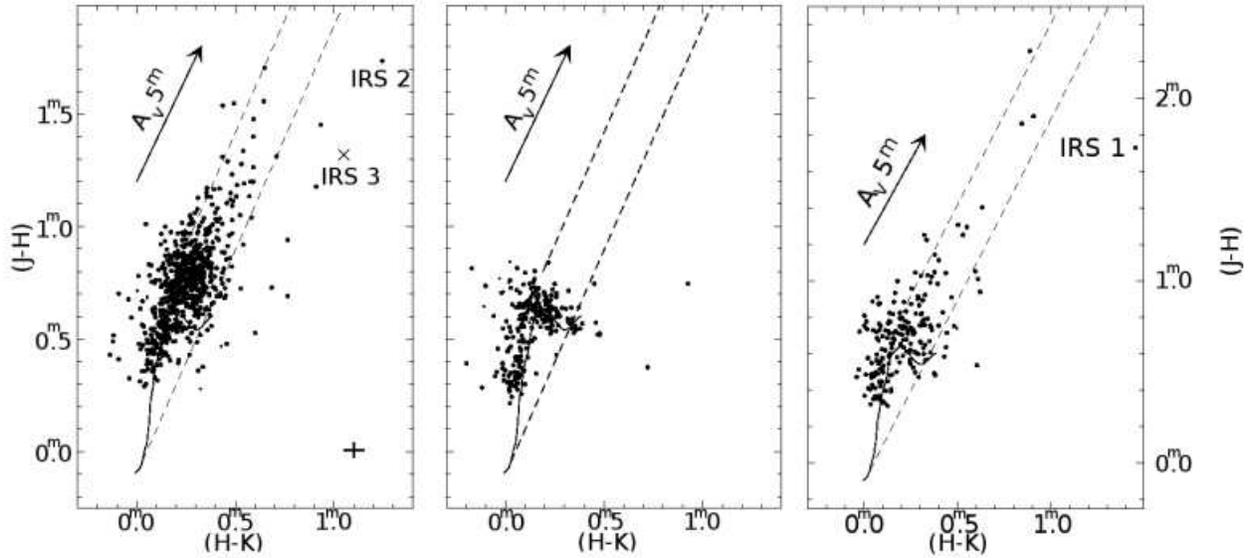}
   \caption{SOFI \J$-$\H, \H$-$\Ks\ colour-colour diagram of CG~1 (left panel), the OFF region (middle) and CG~2 (right). The dashed lines show the effect of the \citet{bessellbrett88} (BB88) reddening law. The unreddened main sequence given in \citetalias{bessellbrett88} is shown with continuous line. A 5\umag\ reddening vector has also been included in the diagram. The average formal error of a data point is indicated in the lower right of the left panel. The discovered IR excess stars have been marked. The cross shows CG~1 IRS~3, which is not included in the final photometry catalogue.}
\label{fig:twocolor}
\end{figure*}

The \J$-$\H, \H$-$\Ks\ colour-colour diagrams of CG~1 and the OFF field are shown in the left and middle panels of Fig. \ref{fig:twocolor}. The \citet[][BB88]{bessellbrett88} main sequence converted into the 2MASS system using the conversion formulae by \citet{carpenter01} is also plotted. The solid lines indicate the locations of unreddened main sequence. The dashed lines are reddening lines drawn using the \citetalias{bessellbrett88} reddening law; normal reddened MS stars lie between or close to these lines. The infrared excess stars should lie right of the rightmost reddening line. The discovered IR excess stars are named. The arrow indicates a reddening vector of 5\fm0. The errorbars indicate the average errors, which are about half of the maximum errors. Several stars are seen to the right of the reddening line in Fig. \ref{fig:twocolor}. In the ON field they are fainter than m$_{\J} \sim 20\fm2$. Only one star brighter than this ($\J-\H=1.74$, $\H-\Ks=1.24$, m$_{\J}=$19\fm6) has infrared excess; this object (CG~1 IRS~2) is discussed further in Sect. \ref{sect:irexcess_cg1}.

The OFF field shows no field-wide reddening effects, thus allowing its use as a reference field when studying the colour excesses. The objects to the right of the reddening line in the OFF field are faint (m$_{\J}>19\fm3$). The low signal-to-noise ratio of these objects (integration time of the sky field is lower than in the ON field) can cause Source Extractor (SE) to classify extended objects as stars. The objects below the reddening line are faint, and at a distance of 300 pc they would be substellar. It is possible that they are long-period variables or extragalactic. None of the objects showing infrared excess in the OFF field are detected in the Wide-field Infrared Survey (WISE) images (see Sect. \ref{sect:starformation} for more discussion on WISE). The objects with negative ($\H-\Ks$) colours are possibly extragalactic. Both the ON and OFF field diagrams show a slight offset ($\H-\Ks$ $\sim$ 0\fm04) between the tabulated main sequence and the data. This offset is the same in both the ON and OFF regions, and thus it does not affect the reddening estimates. For comparison, 2MASS PSC data from a 1000\arcsec\ box around the OFF position has a similar offset between the magnitudes and the tabulated main sequence. This suggests that the stellar colours differ from the tabulated colours at high galactic latitudes, possibly because of metallicity or because the transformation of the \citetalias{bessellbrett88} main sequence to the 2MASS photometric system is not accurate.

\subsubsection{CG~2}
\label{sect:results_cg2ph}

Except for the offset, the CG~2 \J$-$\H, \H$-$\Ks\ colour-colour diagram in the right panel of Fig. \ref{fig:twocolor} shows good agreement with the main sequence and has also several reddened stars. The NIR excess star at $\J-\H=1.73$, $\H-\Ks=1.45$
(CG~2 IRS~1) is located in the head of CG~2, just east of the globule core. The star is discussed in detail in Sect. \ref{sect:irexcess_cg2}. Stars below the reddening line are all fainter than $18\fm7$\ in $m_{J}$, most of them fainter than $19\fm4$.

\section{Discussion} \label{sect:discussion}

\subsection{Visual extinction} \label{sect:visualextinction}

Visual extinction ($A_{V}$) maps were derived from the data using the NICER method introduced in \citet{nicer_lombardietalves}. The method utilises colour excesses from the ON and the unreddened OFF field stars to determine the relative extinction between the two fields. 

The \citetalias{bessellbrett88} extinction law was used to estimate the visual extinction. The \citet{mathis90} and \citet{riekelebofsky85} reddening laws produce $\sim$9\% and $\sim$8\% higher $A_{V}$ values than \citetalias{bessellbrett88}, respectively. However, the relative variation in the extinction within the final map will be the same irrespective of the reddening law used. The $A_{V}$ values in the text have been derived with the \citetalias{bessellbrett88} law.

In the SIRIUS data, the Gaussian used for smoothing has an FWHM of 72\arcsec\ and a pixel scale of 36\arcsec, whereas the SOFI data was smoothed using a 45\arcsec\ FWHM Gaussian and the pixel scale is 22.5\arcsec. The pixel values for any empty pixels were interpolated from the neighbouring pixels. Due to higher resolution and limiting magnitudes, the extinction values in the SOFI map can reach higher values (and sharper peaks) than in the SIRIUS map.

\subsubsection{CG~1} \label{sect:visualextinction_cg1}

\begin{figure}
\resizebox{\hsize}{!}{
\includegraphics[width=17cm] {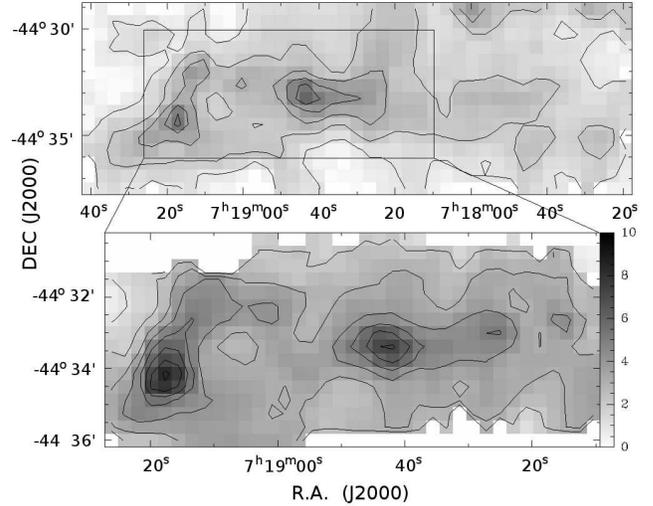}
}
   \caption{Visual extinction map of CG~1. Upper panel: SIRIUS data. The contours are from 1 to 5 magnitudes in steps of 1\umag. Lower panel: SOFI data. The contours are from 1 to 8 magnitudes in steps of 1\umag. The grey scale is the same in both panels.}
\label{fig:cg1avmap}
\end{figure}

The extinction maps of CG~1 derived from the SIRIUS and SOFI data are shown in Fig. \ref{fig:cg1avmap} upper and lower panels, respectively. The lowest contour value and the contour step in the figure is 1\fm0. Typical values for the extinction outside the globule are less than 1\umag\ in the SIRIUS data.
The highest extinction, 8\fm3, is seen in the SOFI map. The large-scale structures in the SIRIUS and SOFI $A_{V}$ maps agree well. Two large clumps are seen in both the SOFI and the SIRIUS maps and will be referred to as the ``Head'' and ``Middle'' clumps in the following. A further weaker clump called ``Tail'' is seen in the SIRIUS $A_{V}$ map outside the area imaged with SOFI. The SOFI map has a small local maximum at the western edge, but this cannot be seen in the SIRIUS map. This feature is most likely an artefact caused by the bright, saturated star at this location in the SOFI image.

The Head maximum peaks west of the YSO and it agrees well with the location and shape of the maximum C$^{18}$O emission \citepalias[see][]{cg1head} in the head of CG~1. Towards north the eastern rim of the extinction in the head follows the sharp edge of the globule seen in Fig. \ref{fig:cg1_3colorsofi}. The Head maximum is thus located very close to the surface of the exposed side of the globule. The position of the Middle maximum agrees well with the extended surface emission seen in the SIRIUS image and the filamentary structure in the SOFI image. There is indication in the SOFI data that the Middle clump is fragmented into two.

Unlike to what is seen in SOFI data, the maximum  $A_{V}$ in CG~1 in the SIRIUS data is not seen in the Head but in the Middle maximum. This happens because of the small size of the $A_{V}$ maximum in the Head is diluted in the large SIRIUS pixels. In addition, because SIRIUS has lower limiting magnitudes than SOFI, the most reddened stars in or near the $A_{V}$ maximum in the SOFI image are not detected in the SIRIUS image. 

The maximum visual extinction in the head computed for the SOFI data with the \citet{mathis90} extinction law differs from the results obtained in \citetalias{cg1head}. This is due to the maps using different binning, which causes the stars in the N1 frame to fall in different pixels. Because the $A_{V}$ value in a single pixel is computed as an average of all the extinctions from the stars in the area of that pixel, the extinction values in the map can change when the pixel centre coordinates are changed. The $A_{V}$ peak in the CG~1 head is sharp, which makes it especially sensitive to a change in the pixel coordinates.

\subsubsection{CG~2} \label{sect:visualextinction_cg2}

\begin{figure}
\resizebox{\hsize}{!}{
\includegraphics{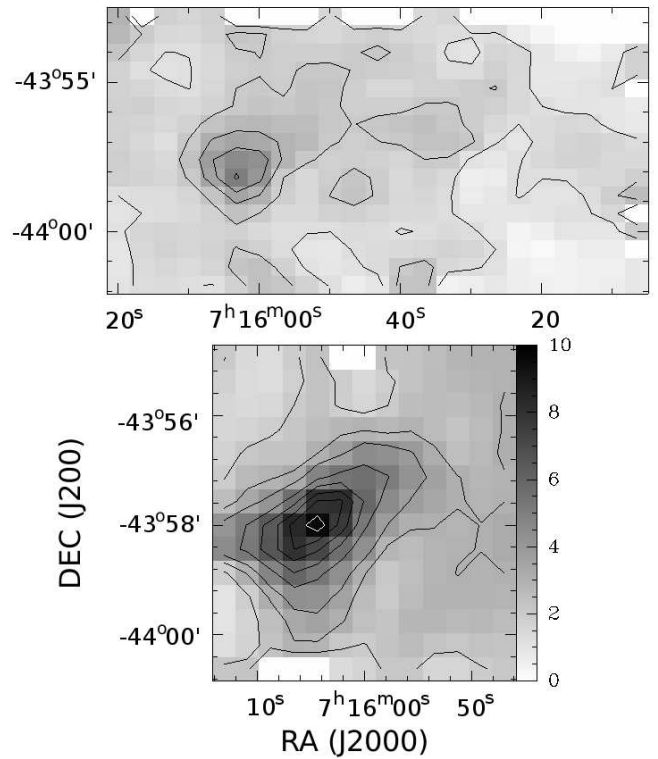}
}
   \caption{Visual extinction map of CG~2. Upper panel: SIRIUS data. The contours are from 1\umag\ to 5\umag\ in steps of 1\umag. Lower panel: SOFI data. The contours are from 2\umag\ to 9\umag\ in steps of 1\umag. The grey scale is the same in both panels. 
}
\label{fig:cg2avmap}
\end{figure}

The SIRIUS and SOFI $A_{V}$ maps of CG~2 are shown in the upper and lower panels of Fig. \ref{fig:cg2avmap}, respectively. The lowest contour levels in the SIRIUS and SOFI maps are 1\fm0 and 2\fm0, respectively. The contour step is 1\fm0. The observed distribution of extinction in CG~2 is similar to the one observed in CG~1. A sharp maximum is seen in the globule head and a weaker maximum lies in the tail. The tail maximum is, however, relatively weaker when compared to the head clump than is the case in CG~1. In the SOFI map the visual extinction follows the filamentary structure that extends from the apex of the globule to the two bright stars in the NW direction. The $A_{V}$ in the head peaks at 5\fm0 in SIRIUS and at 9\fm6 in SOFI. In the tail a maximum of 2\fm5 is observed. Outside the globule the $A_{V}$ varies between 1\fm1 and 1\fm4. 

\subsection{Cloud mass and mass distribution} \label{sect:cloudmass}

The mass of a cloud can be estimated using the \citet{bohlinetal78} relation between the H$_{2}$ column density and visual extinction, $N(H_{2})=0.94*10^{21}$ cm$^{-2}$ mag$^{-1}$, where E(B-V)=3.1 is assumed. The extinction depends linearly on the molecular hydrogen column density, and thus $A_{V}$ maps trace the cloud mass distribution. The choice of a reddening law and the adopted distance to the object will influence the total mass of the cloud, but they do not influence the relative mass distribution within the $A_{V}$ map. Thanks to its better resolution, SOFI data is used to estimate the clump masses. However, the masses estimated from SIRIUS data do not differ significantly when using $A_{V}$ contours that define an area equal in size.

\subsubsection{CG~1} \label{sect:cloudmass_cg1}

The CG~1 total mass as derived from the SOFI map, assuming a distance of 300 pc and using the \citetalias{bessellbrett88} extinction law, is 41.9 \Msun. Using the \citet{riekelebofsky85} reddening law the mass would be 45.7 \Msun. The mass estimate for the Head clump inside the SOFI $A_{V}$ contour of 4.0\umag\ is 6.9 \Msun. In the Middle clump, a contour of 4.0\umag\ yields 5.9 \Msun. The Tail clump mass inside the SIRIUS 2.5\umag\ contour is 2.2 \Msun.

To investigate the M$_{Head}/$M$_{Tot}$ mass ratio, the cloud is divided into two along the right ascension at 7\hour18\minute55\second. Using 2.0\umag\ as the lower limit, the SOFI data gives the mass in the head as 16.8 \Msun\ and the total mass of the imaged portion of CG~1 as 40.6 \Msun. The same magnitude limit for SIRIUS yields 11.3 \Msun\ for the head and 37.1 \Msun\ for the entire imaged area of CG~1. The part of the tail that was imaged only by SIRIUS accounts for a mass of 8.6 \Msun.

SOFI data provides a mass fraction of 0.41 and SIRIUS data 0.31 for the head. The discrepancy can be explained through the ``extra'' SIRIUS frame I4. Even though there is only a single, weak clump in this frame, there is a noticeable portion of mass spread uniformly throughout the frame. If the mass from the ``extra'' SIRIUS frame is ignored, the mass fraction for SIRIUS $>$1\umag\ becomes 0.34. The mass estimates may be uncertain by 50-100\% owing to systematic errors in the Av values and the distance estimate. However, the mass ratio between the head and the tail of the globule contains significantly less uncertainty because it is a differential measurement, and the error in the absolute value of the mass is likely to be fairly uniform throughout the cloud.

\citet{harjuetal90} calculated a total molecular mass of 20--45\Msun\ for CG~1 based on the \thco-\Htwo\ and I$_{CO}$-N(\Htwo) conversion ratios. From the \thco map they determine that 25 \% of the mass is located in the head. This agrees with our results for the mass fraction of the head, particularly when examining the larger SIRIUS-imaged area.

\subsubsection{CG~2}

The CG~2 total mass inside the 1\umag\ contour is 31.0 \Msun\ for the larger SIRIUS $A_{V}$ map and 13.1 \Msun\ for the SOFI map. The SOFI contour 4\umag\ gives a mass of 4.5 \Msun\ for the head. Dividing CG~2 into two along the right ascension at 7\hour15\minute45\second\ and using the 1\umag\ contour, the mass fraction of the head is 0.62.

The IRAS HIRES 12, 25, 60, and 100 $\mum$ images of CG~2 are shown in the online Fig. \ref{fig:cg2_irasall}. The CG~2 total length in the IRAS HIRES images is 28\arcmin, most of which is located outside our NIR frames. There is a local maximum in the tail in the IRAS 25 $\mum$ image at 7\hour15\minute20\second\ but no corresponding feature in the SIRIUS $A_{V}$ map (Fig. \ref{fig:cg2avmap}, upper panel). The IRAS 60 $\mum$ map has a local maximum $\sim$5\arcmin\ from the head, which corresponds well to the local maximum in the SIRIUS $A_{V}$ map. The portion of the tail not imaged in NIR contains one smaller clump, which has IRAS 60 and 100 $\mum$ fluxes less than those in the head. Including this would decrease the mass fraction from the value calculated above.

\subsection{Star formation in CG~1 and CG~2}
\label{sect:starformation}

So far two stars, the pre-main-sequence star NX Pup \citep{reipurth83} and \object{2MASS J07192185-4434551} which is embedded in the dense globule head 70\arcsec\ west of NX Pup \citepalias{cg1head}, are known to have formed in CG~1. NX Pup is actually a close binary, and a third T Tau star lies just 7\arcsec\ NE of it \citep{bernaccaetal93, brandneretal95}. The IRAS point source \object{IRAS 07178-4429} has earlier been often associated with NX Pup. However, its nominal position lies approximately halfway between NX Pup and  2MASS J07192185-4434551. As shown in \citetalias{cg1head}, only the 12 and 25 $\mum$\ emission originates in NX Pup and the 60 and 100 $\mum$ emission in the embedded 2MASS star. 2MASS J07192185-4434551 is referred to as CG1 IRS~1 in the following.

Signs of further star formation in CG~1 and CG~2 were searched for by combining the NIR imaging and the photometry presented in this paper and images from the Wide-field Infrared Survey Explorer (WISE). WISE imaged the sky in four bands: 3.4, 4.6, 12, and 22 $\mum$ \citep{wisepaper}. CG~1 and CG~2 All-Sky Release Catalog data and WISE All-Sky Release images, public since March 2012, were retrieved via IRSA.

The WISE images showed counterparts to three NIR excess objects, two in CG~1 and one in CG~2. None of these have optical counterparts in either the DSS or the ESO and AAO R band images. These new objects are referred to as CG~1 IRS~2, CG~1 IRS~3, and CG~2 IRS~1 in the following. The SOFI and WISE photometry and archival catalogue data (Akari, IRAS, Spitzer if available) retrieved via IRSA for the five stars in CG~1 and CG~2 are listed in Table \ref{table:sed_photometry}. The catalogued flux values were adopted when available, and for Spitzer the fluxes were extracted with MOPEX\footnote{http://irsa.ipac.caltech.edu/data/SPITZER/docs/dataanalysistools/\newline tools/mopex/} as aperture photometry. WISE Catalog data were expressed in magnitudes and converted into Janskys with the zero-point magnitudes listed in the WISE All-Sky Data Release Product Explanatory Supplement\footnote{http://wise2.ipac.caltech.edu/docs/release/allsky/expsup/}.

\subsubsection{NX Puppis}
\label{sect:nxpuppis}

\citet{schoelleretal96} derived the masses $\approx$ 2.0 \Msun\ and $1.6-1.9$ \Msun\ and ages $3-5 * 10^{6}$ and $2-6 * 10^{6}$ years for the binary NX Pup components A and B, respectively. \citet{brandneretal95} give component B a wider range of mass, $1.2-2.5$ \Msun\, and a possibility to be as young as $0.3 * 10^{6}$ yr. The third component 7\farcs 0 away from the binary, NX Pup C, has a mass of 0.30 \Msun. We have not modelled NX Pup as it has already been done by \citet{schoelleretal96} and \citet{brandneretal95} and the binary nature of the system is unresolved in our observations. However we note that the IRAS 60 and 100 $\mum$ fluxes deviate strongly from the modelled binary spectral energy distribution (SED) in \citet{brandneretal95}. This is expected because the mid and far IR signal is not connected with NX Pup at all.

\subsubsection{CG~1 IRS~1 (2MASS J07192185-4434551)}
\label{sect:yso}

This object was discussed in \citetalias{cg1head} where it was identified as a young stellar object. A molecular hydrogen object \object{MHO 1411}, which is a probable obscured Herbig-Haro (HH) object, lies $\sim$ 90\arcsec\ west of CG~1 IRS~1. \jhks\ photometry of CG1 IRS~1 is not possible as it is totally obscured in \J\ and \H\ bands and only reflected light is seen. The central object is visible in \Ks\ but even here most of the light is due to scattering in the circumstellar material. Hence we have excluded these bands from our photometry. The IRAS Point Source Catalog fluxes of IRAS 07178-4429 at 60 and 100 $\mum$ are attributed to CG1 IRS~1 in Table \ref{table:sed_photometry}. Akari observations are both from the FIS Bright Source Catalogue and IRC Point Source Catalogue.

\subsubsection{CG~1 IRS~2 (\object{WISE J071846.52-443252.9})}
\label{sect:irexcess_cg1}

This object is located in the tail part of CG~1, in what is identified as an area of extended surface brightness in Fig. \ref{fig:cg1_3colorsofi}, and at the NE edge of the SOFI Middle clump in Fig. \ref{fig:cg1avmap}. The colours derived from SOFI observations are $\J-\H=1.74$ and $\H-\Ks=1.24$ which put it in the upper right-hand corner of the \J$-$\H, \H$-$\Ks\ colour-colour diagram in the left panel of Fig. \ref{fig:twocolor}. The WISE images show a matching object in all wavelengths.

\subsubsection{CG~1 IRS~3 (\object{WISE J071814.71-443420.0})}
\label{sect:cg1b_star}

This object is detected in the tail of CG~1. It is not included in the final SOFI photometry catalogue because SE classified it as a non-stellar object. WISE also flags it as an extended object, possibly due to confusion noise, since there are several stars located within $\sim$7\arcsec\ of IRS~3 in the NIR images. The highest resolution of WISE is 6\farcs1 at 3.4 $\mum$. In this band WISE shows a bright peak at IRS~3 and at one of these close stars, but the WISE Source Catalog lists only IRS~3. We have included CG~1 IRS~3 in our NIR excess object list because of its location in CG~1 and because it can be seen in the WISE 22 $\mum$ image.

\subsubsection{CG~2 IRS~1 (\object{WISE J071604.50-435742.6})}
\label{sect:irexcess_cg2}

\begin{figure*}
\centering
\includegraphics [width=17cm] {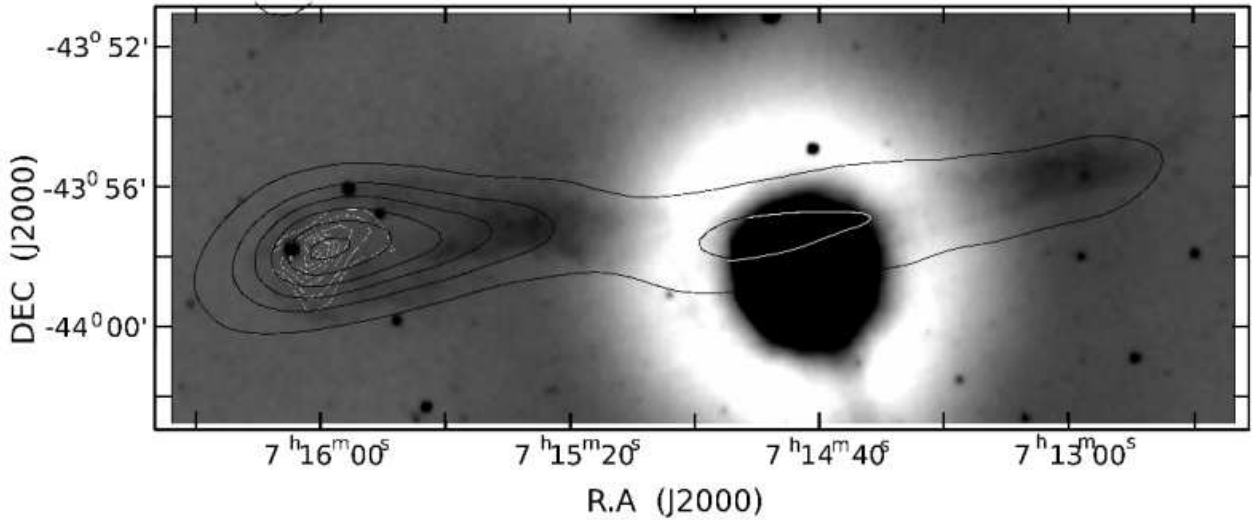}
   \caption{WISE 12$\mum$ image of CG~2. Arbitrary contours of IRAS 100 $\mum$ HIRES processed image (black lines) and the extinction derived from the SOFI data (grey lines) are also shown. The strong artefact in the CG~2 tail is a reflection of a nearby bright star.}
\label{fig:cg2_wise3_100con}
\end{figure*}

This star lies in the head of CG~2 between the globule apex and the region of maximum visual extinction. SOFI photometry gives the NIR colours $\J-\H=1.73$, $\H-\Ks=1.45$. In the \jshks\ images it is clearly red and appears slightly extended. In the 22 $\mum$ WISE image, this is the brightest object in the CG~2 head. WISE 12 $\mum$ image of CG~2 is shown in Fig. \ref{fig:cg2_wise3_100con}. Contours of the HIRES-processed 100 $\mum$ map and of visual extinction derived from the SOFI data are also shown. The large artefact in the CG~2 tail is due to a reflection from a bright nearby star. The IRAS 25 $\mum$ grey scale and IRAS HIRES-processed 60 $\mum$ and 100 $\mum$ contours are shown in the online Fig. \ref{fig:cg2_w4_conwiseiras}.
\citet{bhatt93} associated \object{IRAS 07144-4352} with the head of CG~2. Even though only low-quality IRAS PSC 25 $\mum$ flux is available for the IRAS source, a point source is visible in the HIRES 25 $\mum$ grey scale image. The position of this point source coincides with that of CG~2 IRS~1. The HIRES 100 $\mum$ contours are centred SW of the 25 $\mum$ source and trace the dense globule core (Fig. \ref{fig:cg2_wise3_100con}) rather than IRS~1. Similar to IRAS 07178-4429 in the CG~1 head, which traces emission both from NX Pup and CG~1 IRS~1 \citepalias[see][]{cg1head}, IRAS 07144-4352 also consists of two components: the molecular core and CG~2 IRS~1. The IRAS 60 $\mum$ flux traces, at least partly, CG~2 IRS~1 and the 100 $\mum$ flux the dense globule core.
The Akari FIS catalogue contains a source that has coordinates that are offset $\sim$29\arcsec\ southeast of the WISE coordinates. Neither the IRAS nor the AKARI far infrared fluxes are associated with CG~2 IRS~1 in Table \ref{table:sed_photometry}.

\subsubsection{SED modelling}
\label{sect:sedmodel}

\begin{figure*}
\centering
\includegraphics [width=14cm] {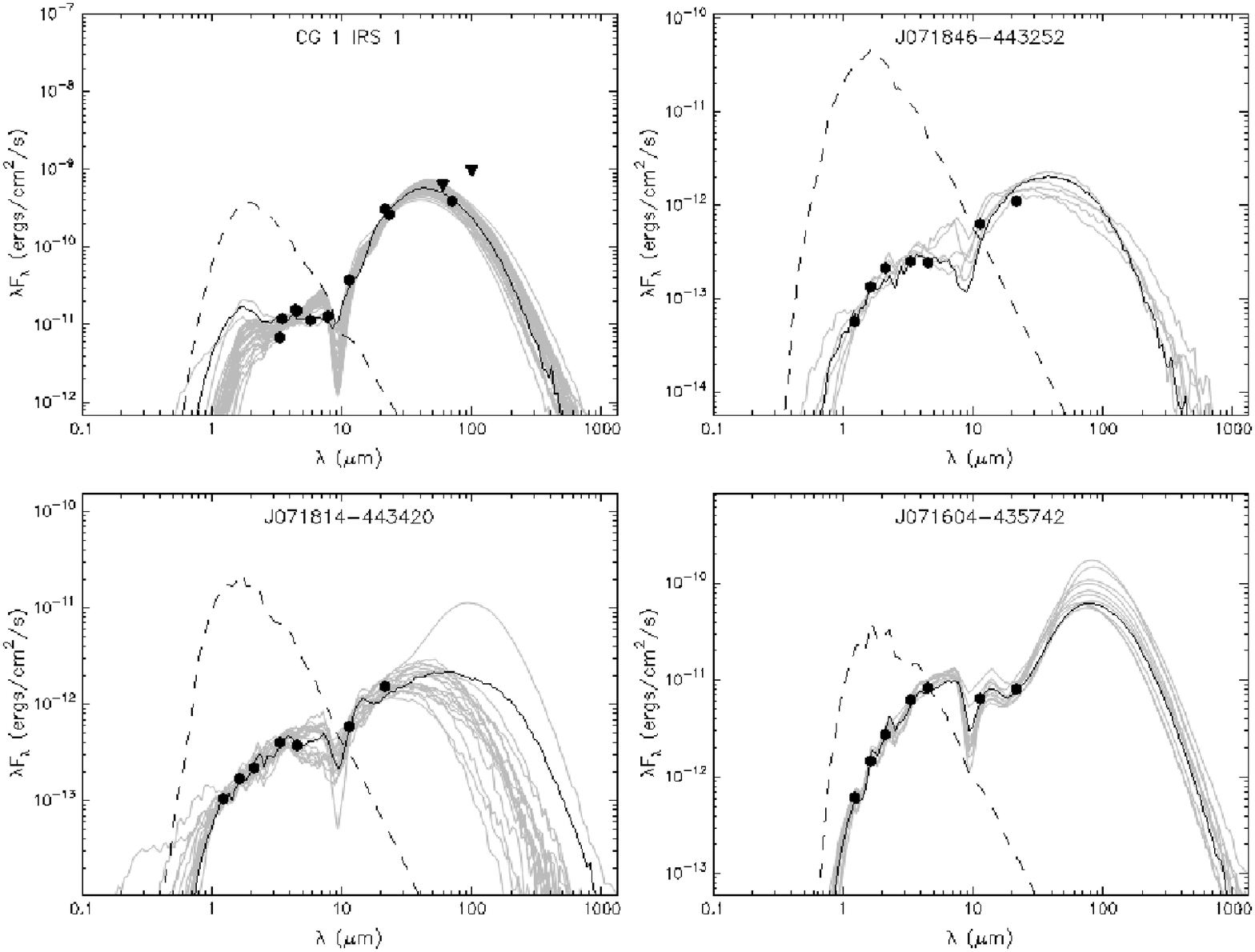}
   \caption{SED fits for our NIR excess sources. NX Pup is omitted as it has been previously modelled (see text). The filled circles are the observed fluxes and the triangles mark the fluxes where there is only an upper limit. The solid line marks the best fit, and the grey lines shows fits with $\chi^{2}-\chi_{best}^{2} < 3$ per each data point. The dashed line indicates the unreddened SED of the stellar photosphere of central source in the best fitting model.}
\label{fig:allsedfits}
\end{figure*}

A SED fitting tool by \citet{robitailleetal07} is available online. The SOFI, WISE, IRAS, and Spitzer data in Table \ref{table:sed_photometry} were used as the input. The errors were set to 10\% although the true errors are smaller than that. The results are only indicative especially because of the lack of FIR and submm data. Only CG~1 IRS~1 has a measured flux at 100 $\mum$, but for the rest of the sources the longest observed wavelength is the WISE 22 $\mum$. The distance to the objects was allowed to vary between 300 and 450 pc to reflect the literature estimates for CG~1. For CG~1 IRS~3, a range of 200--600 pc was allowed because of its location far in the CG~1 tail. The interstellar extinction was allowed to vary from 0.1\umag\ to a maximum estimated by using the $A_{V}$ maps.
The age, mass, temperature, and luminosity of the central source, $A_{V}$, distance and the $\chi^{2}$ for the best fit are listed in Table \ref{table:sedfit_results}. The mass and age ranges for all good fits (i.e. fits that have $\chi^{2}-\chi_{best}^{2} < 3$ per each data point) are also listed. The SED plots with the fits are shown in Fig. \ref{fig:allsedfits}. The uncertainty of the long wavelength part of the SED due to the lack of data points is visible in the plots. Online Fig. \ref{fig:allsedfitsagemass} shows the age-mass relation derived from the good fits. All fits point to low-mass stars.

\begin{table*}
\caption{The photometry for NX Pup and the new NIR excess candidates in CG~1 and CG~2.}
\label{table:sed_photometry}
\centering
\begin{tabular}{c c c c c c}
\hline\hline
 & NX Pup & CG~1 IRS~1 & CG~1 IRS~2 & CG~1 IRS~3 & CG~2 IRS~1 \\
\hline
J\tablefootmark{a} & 590 (8.579) & \ldots & 0.0235 (19.58) & 0.0435 (18.91) & 0.25 (17.01) \\  
H\tablefootmark{a} & 1250 (7.825) & \ldots & 0.0749 (17.84) & 0.0943 (17.59) & 0.791 (15.28) \\ 
K\tablefootmark{a} & 2470 (6.080) & \ldots & 0.153 (16.60) & 0.158 (16.56) & 1.96 (13.83) \\
3.4\tablefootmark{b} & 5184.62 (4.440) & 7.75 (11.504) & 0.281 (15.104) & 0.453 (14.586) & 6.923 (11.626) \\ 
3.6\tablefootmark{c} & \ldots & 13.94 & \ldots & \ldots & \ldots \\
4.5\tablefootmark{c} & \ldots & 23.39 & \ldots & \ldots & \ldots \\
4.6\tablefootmark{b} & 8593.90 (3.252) & 23.05 (9.681) & 0.369 (14.170) & 0.579 (13.680) & 12.853 (10.315) \\ 
5.8\tablefootmark{c} & \ldots & 22.29 & \ldots & \ldots & \ldots \\
8.0\tablefootmark{c} & \ldots & 34.24 & \ldots & \ldots & \ldots \\
9.0\tablefootmark{e} & 5197 & \ldots & \ldots & \ldots & \ldots \\
12\tablefootmark{b} & 5047.81 (1.994) & 144.91 (5.849) & 2.403 (10.300) & 2.271 (10.361) & 24.769 (7.767) \\ 
12\tablefootmark{d} & 6682 & \ldots & \ldots & \ldots & 250\tablefootmark{h} \\ 
18\tablefootmark{e} & 4652 & 1029 & \ldots &  \ldots & \ldots \\
22\tablefootmark{b} & 4816.84 (0.599) & 2236 (1.432) & 8.279 (7.511) & 11.261 (7.177) & 58.879 (5.381) \\ 
24\tablefootmark{f} & \ldots & 2092 & \ldots & \ldots & \ldots \\
25\tablefootmark{d} & 7604 & \ldots & \ldots & \ldots & 370\tablefootmark{h} \\ 
60\tablefootmark{d} & \ldots & 13120 & \ldots & \ldots & 440\tablefootmark{h} \\ 
65\tablefootmark{g} & \ldots & 7697 & \ldots & \ldots & \ldots \\
70\tablefootmark{f} & \ldots & 9314 & \ldots & \ldots & \ldots \\
90\tablefootmark{g} & \ldots & 9672 & \ldots & \ldots & 656 \\
100\tablefootmark{d} & \ldots & 33590 & \ldots &  \ldots & 8880\tablefootmark{h} \\ 
140\tablefootmark{g} & \ldots & 17370 & \ldots & \ldots & 6259 \\
160\tablefootmark{g} & \ldots & 15570 & \ldots & \ldots & \ldots \\
\hline
\end{tabular}
\tablefoot{The flux unit is mJy. For JHK and WISE we also list the magnitude in parenthesis.
\tablefoottext{a}{SOFI.}
\tablefoottext{b}{WISE.}
\tablefoottext{c}{Spitzer IRAC.}
\tablefoottext{d}{IRAS.}
\tablefoottext{e}{Akari IRC.}
\tablefoottext{f}{Spitzer MIPS.}
\tablefoottext{g}{Akari FIS.}
\tablefoottext{h}{Data for the offset IRAS source 07144-4352, flagged U (upper limit) for 12 and 25 $\mum$, B and A for 60 and 100$\mum$, respectively. Used to check the SED model behaviour.}
}
\end{table*}

\begin{table*}
\caption{The results from the best stellar model fits, where mass and age ranges include all models with $\chi^{2}-\chi_{best}^{2} < 3$ per each data point.}
\label{table:sedfit_results}
\centering
\begin{tabular}{c c c c c c}
\hline\hline
 & CG~1 IRS~1 & CG~1 IRS~2 & CG~1 IRS~3 & CG~2 IRS~1 \\
\hline
Mass (\Msun) & 2.84 & 0.50 & 0.21 & 0.14 \\
Mass (\Msun) & 0.24-2.84 & 0.22-1.22 & 0.10-1.49 & 0.11-0.29 \\
Age (yr) & 4.85(5) & 2.15(6) & 4.11(6) & 1.37(5) \\
Age (yr) & 2.8kyr-485kyr & 1.9-7.2Myr & 0.1-9.5Myr & 1.2kyr-140kyr \\
Temperature (K) & 4774 & 3759 & 3234 & 2962 \\
L$_{TOT}$ (\Lsun) & 1.62(1) & 4.05(-1) & 8.98(-2) & 3.65(-1) \\
A$_{V}$ & 10.00 & 3.59 & 4.25 & 6.70 \\
Distance (pc) & 331 & 331 & 209 & 302 \\
$\chi_{tot}^{2}$ & 41.1 & 33.6 & 10.7 & 9.6 \\
\hline
\end{tabular}
\end{table*}

The SED for IRS~1 in the head of CG~1 has a shape of a class I YSO. The best fit gives a mass of 2.8 \Msun\ and an age of 0.5 Myr, which would make IRS~1 more massive than the individual NX Pup components. The other good fits have masses below 1 \Msun, and their ages are in the range of $3-217$ kyr. The age range covers two orders of magnitude, but even with the highest age estimate IRS~1 is a second-generation star compared to NX Pup. The star formation efficiency in the head of CG~1 is about 25\% when using the \citet{schoelleretal96} mass estimates for the AB components of NX Pup, the SOFI $M_{Head}$ estimate and the mean IRS~1 mass from the SED fits (0.55 \Msun).

The NIR excess objects CG~1 {IRS~2 and IRS~3 have similar SEDs and are best fitted with YSOs with ages in the range $\sim$1--7 Myr and masses in the range $\sim$ 0.2--1.3 \Msun. The age is similar to the age of NX Puppis.

In CG~2, the stellar age in the fits ranges from 1.2 kyr to 0.14 Myr and the masses from  0.11 to 0.29 \Msun. The age of this object is comparable to the age of IRS~1. The lack of a large amount of obscuring circumstellar material in CG~2 IRS~1 suggests that this object is more evolved than CG~1 IRS~1.

\subsection{Morphology} \label{sect:morphology}

In addition to the new \jshks\ observations, satellite data (IRAS, Spitzer, WISE) and molecular line data from \citetalias{cg1head} and \citet{harjuetal90} are used to study the structure of CG~1. Only NIR, IRAS, and WISE data are available for CG~2. The large-scale NIR structure of CG~1 and CG~2 has been discussed in Sects. \ref{sect:results_cg1im} and \ref{sect:visualextinction_cg1}.

Large polycyclic aromatic hydrocarbon (PAH) molecules emit strong emission lines at 3.3, 6.2, 7.7, 8.6, 11.2, 12.7, and 16.4 $\mum$ \citep{pahreview_tielens}. PAH emission is thus observed in Spitzer IRAC bands 3.6, 5.8, and 8.0 $\mum$, and in WISE bands 3.4 and 12 $\mum$. The amount of PAH emission in the diffuse interstellar medium observed in the Spitzer IRAC bands has been investigated by \citet{flageyetal06}. Very small grains (VSGs) emit thermally in the mid infrared (MIR) and are expected to be seen only in the Spitzer MIPS 24 and 70 $\mum$ and IRAS 25, 60, and 100 $\mum$ filters. Thermal emission from the large, cold grains is expected only in the IRAS 100 $\mum$ filter but even here a large fraction of the observed emission is due to small grains. Also molecular hydrogen has numerous rotational lines in the near- and mid-infrared. The high spatial resolution of the Spitzer and WISE images allows us to study the detailed structure of CG~1 and CG~2 and especially to estimate whether the structures are due to PAH emission or not.

\subsubsection{CG~1}

The SOFI \jshks\ and \Htwo\ (1--0) S(1) emission structure of the CG~1 head was discussed in \citetalias{cg1head}. The individual Spitzer IRAC images at 3.6, 4.5, 5.8, and 8.0 $\mum$ and MIPS images at 24 and 70 $\mum$ are shown in the online Figs. \ref{fig:cg1_iracall} and \ref{fig:cg1_spitzermips}, respectively. The Spitzer images cover only the head of CG~1. The WISE images of CG~1 at 3.4, 4.6, 12, and 22 $\mum$ are shown in the online Fig. \ref{fig:cg1wise_all}. 
A false-colour WISE image is shown in Fig. \ref{fig:cg1wise3color}. 
The location of the most notable features seen in the head of CG~1 in WISE and Spitzer are marked in the online Fig. \ref{fig:cg1_structure}.
Two cone-like structures symmetrically around CG~1 IRS~1 are seen in the IRAC images, with a bright streamer that originates in CG~1 IRS~1 and falls on the axis of the SW-pointing cone. This feature is seen best at wavelengths longer than 5.8 $\mum$. Below NX Puppis is a bright reflection nebula that is seen only at 3.6 and 70 $\mum$. Surface brightness maxima are observed in the tip of the SW streamer and NW of CG~1 IRS~1 and are seen in the MIPS 24 $\mum$ image. The NW patch is marked with an ellipse. A solid white circle marks MHO~1411.

{\bf The cones:} The spatial resolution of the WISE images is insufficient to resolve the thin filaments originating from CG~1 IRS~1 seen in the SOFI \jshks\ images \citepalias[][Fig. 3]{cg1head}. However, they are faintly visible in the IRAC 3.4 and 4.5 $\mum$ images where another SW pointing filament is seen. These filaments form two cones, which are best seen in the IRAC 3.4, 4.5, and 8.0 $\mum$ colour composite image shown in online Fig. \ref {fig:cg1_spitzer_color}. The cones open to east and SW from CG~1 IRS~1.
The opening angle of the cone to the east is larger than that of the one opening to SW. %WISE\_SW streamer lies on the axis of the SW cone.

{\bf WISE\_SW streamer:} 
The bright, large-scale streamer reaching SW from CG~1 IRS~1 inside the SW cone will be called WISE\_SW. This feature is strong in the WISE 12 and 22 and IRAC 8.0 and MIPS 24 $\mum$ images. It is also faintly visible at 3.4 and clearly detected in the 5.8 $\mum$ image. The maximum WISE\_SW surface emission seen in the 24 $\mum$ image in the tip of the streamer lies just outside the other Spitzer images but is detected in the WISE 3.4, 12, and 22 $\mum$ images. It is also seen in the HIRES-enhanced IRAS 60 $\mum$ image shown in the online Fig. \ref{fig:cg1_irasall} (bottom left). Contour plots of WISE 3.4 and 22 $\mum$ images overlaid on the SOFI \Js\ image are shown in the online Fig. \ref{fig:cg1_js_w1w4con}. 
The contours delineating WISE\_SW in Fig. \ref{fig:cg1_js_w1w4con} are above and clearly offset from the filament seen in \Js. WISE\_SW and the bright filament seen in the \jhks\ imaging are thus different features. The surface brightness maximum in the western tip of WISE\_SW has a corresponding patch on the edge of the globule head NW from CG~1 IRS~1. The NW patch has a lower surface brightness but is visible in the same images as the WISE\_SW streamer except for MIPS 24 $\mum$ where the patch is outside the image. It is suggested that the WISE\_SW streamer and the patch are due to PAH emission in the WISE 12 $\mum$ band. The emission seen in the IRAS HIRES 60 $\mum$ image would be due to thermal emission from very small grains. The source of excitation of the PAH emission and the heating mechanism of the VSGs is not known.

{\bf MHO 1411:} This faint molecular hydrogen object was first reported in \citetalias{cg1head}. A faint object is also seen in the IRAC 3.6 and 4.5 $\mum$ and in the WISE 4.6 $\mum$ image. Its existence has thus been independently confirmed.

{\bf Reflection nebulosity below NX Pup:} 
The bright surface emission coincides with a small, localized component in the CG~1 \ceo mapping reported in \citetalias{cg1head} where the feature was referred to as \ceoSE. The nebulosity can be seen in the IRAC 3.6 $\mum$ image (online Fig. \ref{fig:cg1_iracall}, top left) and the MIPS 70 $\mum$ image (online Fig. \ref{fig:cg1_spitzermips}, right panel). The IRAC data strengthens the claim the nebulosity {in 3.6 $\mum$} is due to NIR scattering of radiation coming from NX Pup. The surface brightness observed at 70 $\mum$ is caused by the thermal emission of very small grains.

{\bf CG~1 Tail:} The large-scale structure of the tail as seen in the NIR extinction and CO emission \citep{harjuetal90} agrees well with the tail seen in the optical images. A new structure is seen in the WISE 3.4, 12, and 22 $\mum$ images (online Fig. \ref{fig:cg1wise_all}) and in the HIRES-enhanced IRAS maps (online Fig. \ref{fig:cg1_irasall}). Extended surface brightness at the southern edge of the tail starts about 13\arcmin\ west of the head and extends $\sim$10\arcmin\ west.
This WISE surface brightness is not seen in the NIR images. In the following this feature is referred to as Tail\_South. The infrared excess star CG~1 IRS~3 lies at the very eastern edge of Tail\_South.

Tail\_South resembles a cometary globule superposed on CG~1. However, the lack of both observed extinction and of strong CO emission \citep{harjuetal90} suggests that it is not a second cometary globule. More likely Tail\_South is emission from PAHs and/or VSGs. IRAS HIRES data shows a local maximum at the position of Tail\_South at 12, 25, and 60 $\mum$. In the WISE images Tail\_South is strongest at 12 $\mum$ and weakest at 3.4 $\mum$. These two wavelengths could be explained with PAH emission, but the emission observed at 22 $\mum$ and in the IRAS HIRES 25 and 60 $\mum$ images requires VSGs. 
This raises another question: What is the origin of excitation of the PAHs and the heating for the VSGs? It cannot be the general interstellar radiation field (ISRF) from the Galactic plane as Tail\_South lies on the shadow side of the CG~1 tail with respect to the Galactic plane. In addition, the head of CG~1 shadows Tail\_South from the emission from NX Pup and the O star $\zeta$ Pup in the centre of the Gum Nebula. However, the multiple system $\gamma^{2}$ Vel in the central part of the Gum Nebula is located so that a line drawn from the system through the tip of WISE\_SW would follow the NE edge of Tail\_South. In this view the SW streamer shields the part of the CG~1 tail that is closest to the head and leaves the rest of the tail exposed to the UV radiation of $\gamma^{2}$ Vel. 
In the tail, the IRAS 12 and 25 $\mum$\ emission peak is separated from the emission in 60 and 100 $\mum$. Similar behaviour has been detected e.g. in L1780 by \citet{ridderstadetal06}. They suggest that this indicates two spatially segregated dust populations.

\begin{figure*}
\centering
\includegraphics [width=17cm] {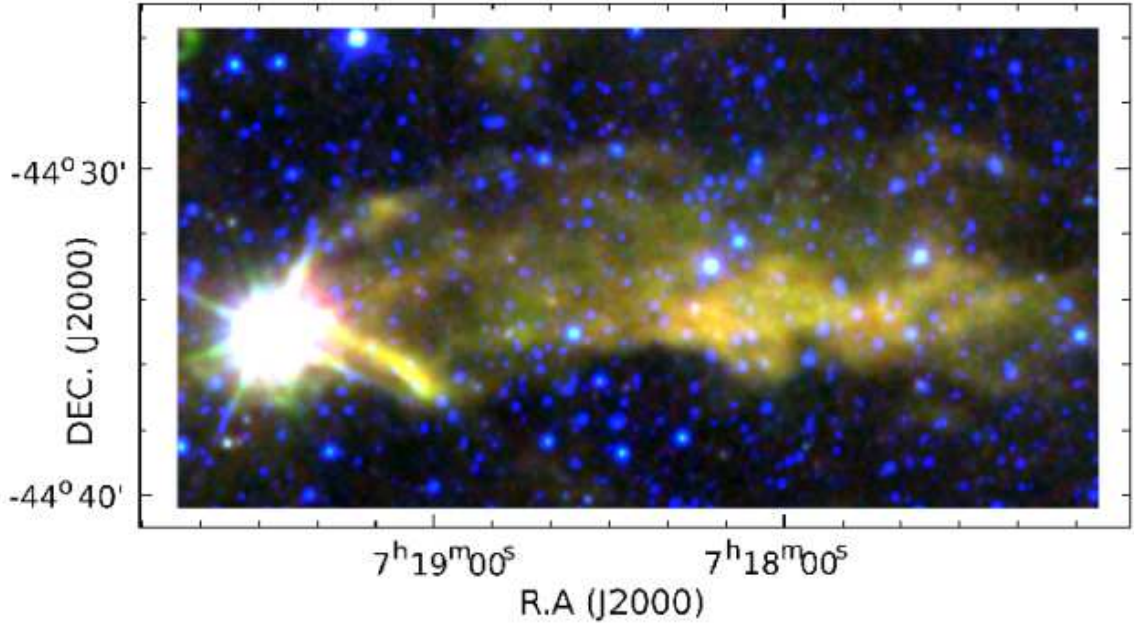}
   \caption{WISE CG~1 false-colour image. The 4.6, 12, and 22 $\mum$ bands are coded in blue, green, and red, respectively.}
\label{fig:cg1wise3color}
\end{figure*}

\subsubsection{Does CG~1 IRS~1 drive an outflow?}
\label{sect:outflow}

CG~1 IRS~1 has the appearance of a Class I YSO (see Fig. \ref{fig:allsedfits} and \citetalias{cg1head}). Such objects typically drive molecular outflows and jets. However, in this case there is very little evidence of an outflow in the available molecular line data presented in \citet{harjuetal90}. An indication of a very modest outflow wing is seen in the \twco (3--2) spectra covering the immediate surroundings of CG~1 IRS~1 \citepalias{cg1head}. The only firm indicator of an outflow is MHO~1411, which lies about 90'' west of CG~1 IRS~1. In addition to the molecular hydrogen object, the \jhks\ and Spitzer IRAC 3.6 and 4.5 $\mum$ images reveal two cones that could delineate the walls of a cavity carved by an outflow. 
The cone edges are visible in the IRAC 4.5 $\mum$ image and are thus not due to PAH emission, leaving scattering as the viable option. IRAC observations around HH~46/47 presented in \citet{velusamyetal07} show that the outflow cavity edges can be visible in the IRAC images. The WISE\_SW streamer is puzzling because it lies between the possible cavity walls. 
At 12 $\mum$ it could be due to strong PAH emission  but the detection at 22 and 60 $\mum$ suggests that it is also associated with warm VSGs. Considering the new information provided by Spitzer imaging, it is suggested that CG~1 IRS~1 is driving an outflow, and the outflow cavity walls are seen in reflected light. However, the nature of WISE\_SW and the very weak CO outflow need to be investigated further in detail.

\subsubsection{CG~2}

CG~2 WISE images at 3.4, 4.6, 12, and 22 $\mum$ are shown in online Fig. \ref{fig:cg2_wiseall}. The artefact in the centre of all images is caused by a very bright field star south of the field. The globule head is faintly detected at 3.4 $\mum$ and hardly visible at 4.6 $\mum$. At 12 and 22 $\mum$ the surface brightness is stronger and coincides approximately with the extended surface brightness of the SIRIUS and SOFI images (Figs. \ref{fig:cg2_3colorsirius} and \ref{fig:cg2_3colorsofi}). The NIR excess object CG~2 IRS~1 is the brightest object in the CG~2 head at 22 $\mum$. The narrow tail is mostly covered by the artefact, but a lane of emission is also seen both to east and west of it. This emission lane can be seen in IRAS HIRES contours (Fig. \ref{fig:cg2_irasall}) at 25, 60, and 100 $\mum$.
The WISE artefact covers the western maximum emission at 60 and 100 $\mum$ (Fig. \ref{fig:cg2_wise3_100con}). The total length of CG~2, from the head to the end of the emission lane is $\sim$28\arcmin. 

No information on the extinction in the tail is available because it is not included in the NIR observations. The tail is separated into individual clumps in the HIRES images, but this may possibly be because the HIRES maximum entropy algorithm tends to form compact structures. At the moment it is not possible to say whether the MIR and FIR emission seen in the globule tail is only due to warm small dust particles or if it is because of a combination of both very small and larger sized particles. The extension of the NIR imaging, molecular line observations or submm dust continuum observations are needed to solve this question.

\subsection{CG formation mechanism}

The mass distribution in the shocked cloud is discussed in the literature mostly on a qualitative level, but some estimates have been given. \citet{heathcotebrand83} find that in the SN shock case the fraction $F$ of mass that goes into the tail depends on the velocity of the flow of the gas behind the external shock.
For a typical case they find $F \sim 0.5$. In the RDI simulations of \citet{leflochlazareff94}, 80\% of the mass remains in the head after the initial collapse phase. The cometary phase begins after the collapse phase and lasts 90\% of the estimated 3 Myr lifetime of the CG. The flux of the ionizing radiation can also affect the mass distribution in RDI \citep{gritschnederetal09, bisbasetal11}.

The mass fraction of the CG~1 head has a lower limit of $\sim$0.4 from the SOFI observations and $\sim$0.3 from the SIRIUS. Star formation in CG~1 has taken place first in the globule head with NX Puppis as the first generation and CG~1 IRS~1 the second, triggered generation. The position of these stars near the globule's front edge is in accordance with the RDI-induced formation \citep{sugitanietal89, leechen07, bisbasetal11}. Taking also the masses of these stars into account, the fraction of mass in the CG~1 head is raised by about 7--10\%. As the star formation efficiency (SFE) is typically of the order $\sim$few \% in dark clouds, the original mass of CG~1 head must have been higher before star formation took place. This pushes the mass fraction towards the values from the RDI model. Even though the CG~1 head-to-tail mass ratio may agree with the RDI formation scenario, the disrupting effects of star formation prevent firm conclusions. The globule is in too evolved a stage to reflect the original mass distribution after the triggering event.

The observed head-to-total mass ratio in CG~2 is 0.6. This ratio must have been higher before the formation of CG~2 IRS~1 in the head. This would be a strong argument in favour of RDI being the globule formation mechanism. However, the observed area does not cover the whole tail. The appearance of the tail is by far not as extended as in CG~1. Further observations of the tail are needed before a definitive value can be given for the head-to-tail mass ratio in CG~2.

\section{Summary and conclusions} \label{sect:conclusions}

We have analysed NIR imaging and photometry and available archival data to study the detailed structure and star formation in the cometary globules CG~1 and CG~2 located in the Gum Nebula. The NIR observations were done in the NIR \J, \H, and \Ks\ bands with SOFI at the NTT telescope and SIRIUS at IRSF. In addition, data from IRAS, WISE, Spitzer, and AKARI telescopes were retrieved from online archives.

\begin{itemize}
\item We find two new NIR excess objects in CG~1, and one in CG~2. CG~1 IRS~2 and CG~1 IRS~3 are located in the tail, whereas the previously known young stellar CG~1 IRS~1 is located in the head of the globule \citepalias{cg1head}. The formation of the pre-main sequence star NX Pup in CG~1 is already known \citep{reipurth83}. The object CG~2 IRS~1 located in the head of the globule is the first indication of star formation taking place in CG~2. 
\item According to SED fitting, CG~1 IRS~1 is a class I source. The SED fits to the rest of the NIR excess sources indicate that they are also young, low-mass stars. The spread of the fitted ages for each NIR excess object is, however, large. This is largely due to lack of FIR and submm photometry data for all objects except CG~1 IRS~1.
\item CG~1 IRS~1 in the head of CG~1 is a second-generation star and is indicative of triggered star formation.
\item CG~1 IRS~1 is likely to drive a bipolar outflow. Structures similar to outflow cones are detected in the imaging. The wider eastern lobe opens towards NX Puppis and the western lobe opens to SW. Mysteriously, no strong outflow features are seen in the available molecular data,
but a bright streamer is seen on the axis of the SW cone in MIR.
\item WISE images and IRAS HIRES-processed images suggest the presence of excited PAH particles and warm, very small grains in the CG~1 tail.
\item CG~1 is in far too evolved a stage to reflect the globule mass distribution at the time of its formation. Thus no firm conclusion on the original tail-to-head mass ratio can be obtained. However, when considering the star formation history in the globule, the mass distribution of CG~1 is indicative of RDI triggered formation. 
\item The observed tail-to-head mass ratio in CG~2 is in accordance with the RDI triggered formation. However, since the observations do not cover the tail in whole, further observations are needed before a final conclusion can be drawn.
\end{itemize}

\begin{acknowledgements}
       This work was supported by the Academy of Finland under grant 118653 132291. M.M. acknowledges the support from the University of Helsinki Senat's graduate study grant and from the Vilho, Yrj\"o and Kalle V\"aisal\"a Fund.
 This publication makes use of data products from the Wide-field Infrared Survey Explorer, which is a joint project of the University of California, Los Angeles, and the Jet Propulsion Laboratory/California Institute of Technology, funded by the National Aeronautics and Space Administration. This research has made use of the SIMBAD database, operated at the CDS, Strasbourg, France.
\end{acknowledgements}

\bibliographystyle{aa}
\bibliography{cg1_refarticles.bib}

\Online

\begin{appendix}

\section{Data reduction}
\label{app:datareduction}

The SOFI and SIRIUS data were reduced in the same manner as in \citetalias{cg1head}. IRAF\footnote{IRAF is distributed by the National Optical Astronomy Observatories, which are operated by the Association of Universities for Research in Astronomy, Inc., under cooperative agreement with the National Science Foundation.} and the external XDIMSUM package was utilised. 
For SOFI, each frame was corrected for cross talk. Then bad pixels were masked and cosmic rays removed in all frames. Sky-subtraction was done using the two temporally closest frames. An object mask was created for each frame to mask the field stars. Running sky-subtraction with the object masks also created hole masks for each frame. The flat field and illumination correction files available at the ESO SOFI website were used on the SOFI sky-subtracted images. Evening sky flats were used to flat-field the SIRIUS data. The bad pixel mask, cosmic ray masks, and hole masks were combined to create rejection masks, which were used when co-adding the images. The coordinates were tied to the 2MASS Point Source Catalog (PSC).

For each colour the objects were extracted from the registered frames using Source Extractor v. 2.5.0 \citep[SE,][]{SExtractor}. SE fits an aperture to each object, and this is used to extract the flux information. The SIRIUS photometry was calibrated using 2MASS PSC. The SE photometry of each colour and field was calibrated using 10--14 2MASS PSC stars per field by linearly scaling the frames to the PSC star level. The difference in the magnitudes of the observed field and the PSC was $< \pm$0.01\umag\ in all colours. SOFI-magnitude zero points were fixed using standard star observations. The SOFI magnitudes extracted by SE were converted into the Persson system (1998) and from there to 2MASS magnitudes, as described by \citet{ascensoetal07}. The SOFI magnitude scale was checked using stars in common with the 2MASS PSC.

Several experimentally found selection rules were used to discard galaxies and stars with poor photometric data from the catalogues produced by SE. SE photometry of SIRIUS images were filtered as follows. Objects were removed if they were closer than 80 pixels to the frame border, SE flagged their flux data, their elongation was more than 1.3 or if any colour had a photometric error calculated by SE of larger than 0\fm15. SE also provides keyword CLASS, which assigns a number between 0.0 and 1.0 to describe how star-like the object is (1.0 for stars, 0.0 for galaxies) in each colour. This was utilised so that objects that had a summed-up star index less than 1.6 were discarded. 
The selection rules to exclude non-stellar objects and stars with doubtful photometry in the SOFI SE photometry were the objects closer than 30 pixels (0 for CG~2) of the edge of the registered frame, those SE flagged to have bad flux data, objects with an elongation larger than 1.3 (1.25 for CG~2), or a magnitude error over 0\fm15 in any of the three bands. Objects whose summed-up star indices were less than 2.70 were left out. After a visual inspection, any objects with two unresolved stars in the extraction aperture were removed. The OFF field SE photometry was filtered the same way. In practice the formal error of the remaining objects is less than 0\fm1 in all colours.

\section{Online data}

\begin{figure*}
\centering
\includegraphics [width=17cm] {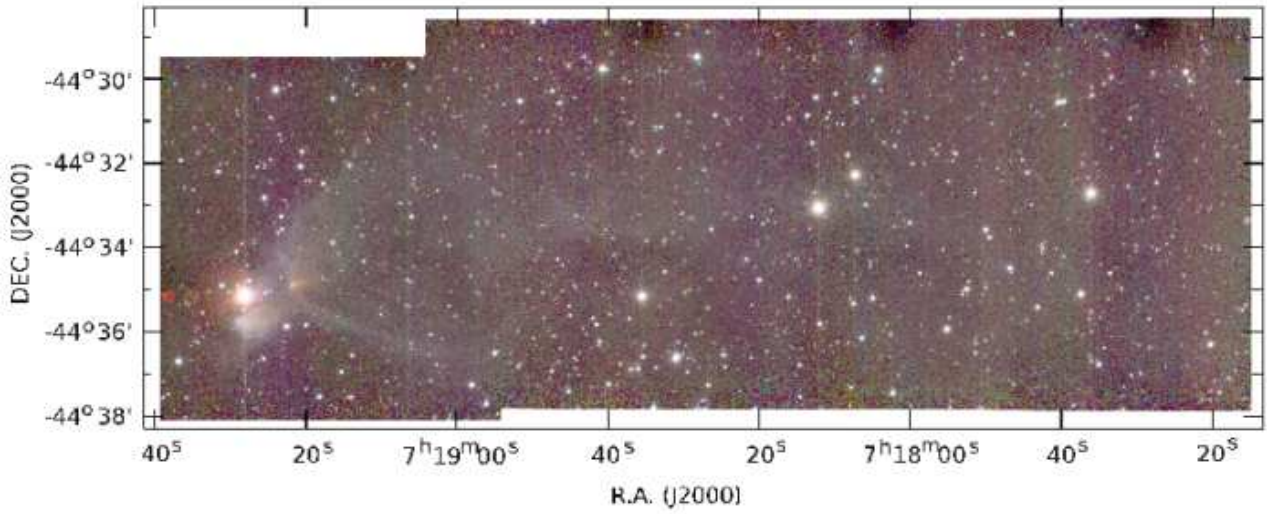}
   \caption{Colour-coded SIRIUS image of CG~1. The \J, \H, and \Ks\ bands are coded in blue, green, and red.}
\label{fig:cg1_3colorsirius}
\end{figure*}

\begin{figure*}
\centering
\includegraphics [width=17cm] {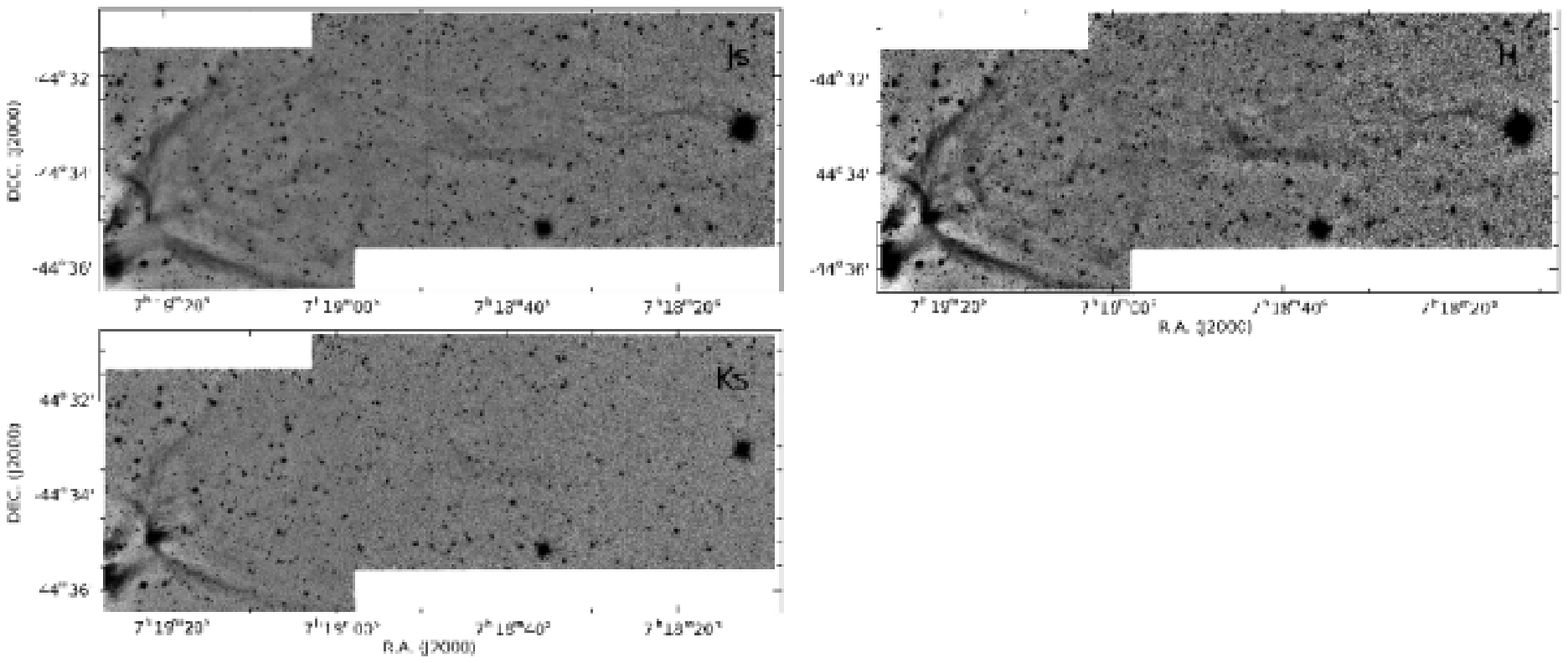}
   \caption{SOFI \Js, \H, and \Ks\ band images of CG~1.}
\label{fig:cg1_sofiall}
\end{figure*}

\begin{figure*}
\centering
\includegraphics [width=17cm] {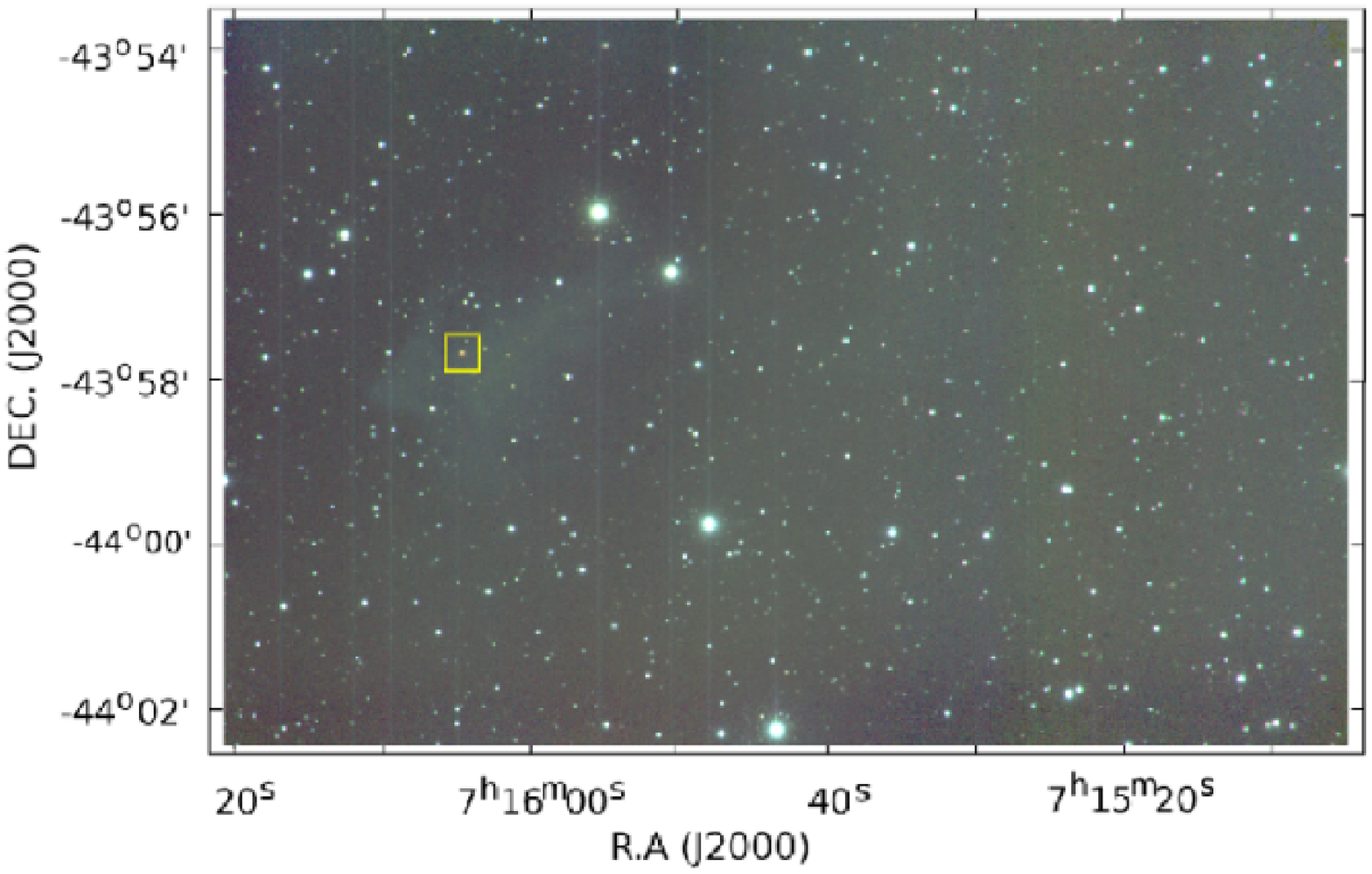}
   \caption{Colour-coded SIRIUS images of CG~2. The \J, \H, and \Ks\ bands are coded in blue, green, and red. The object CG~2 IRS~1 is indicated with a box.}
\label{fig:cg2_3colorsirius}
\end{figure*}

\begin{figure*}
\centering
\includegraphics [width=17cm] {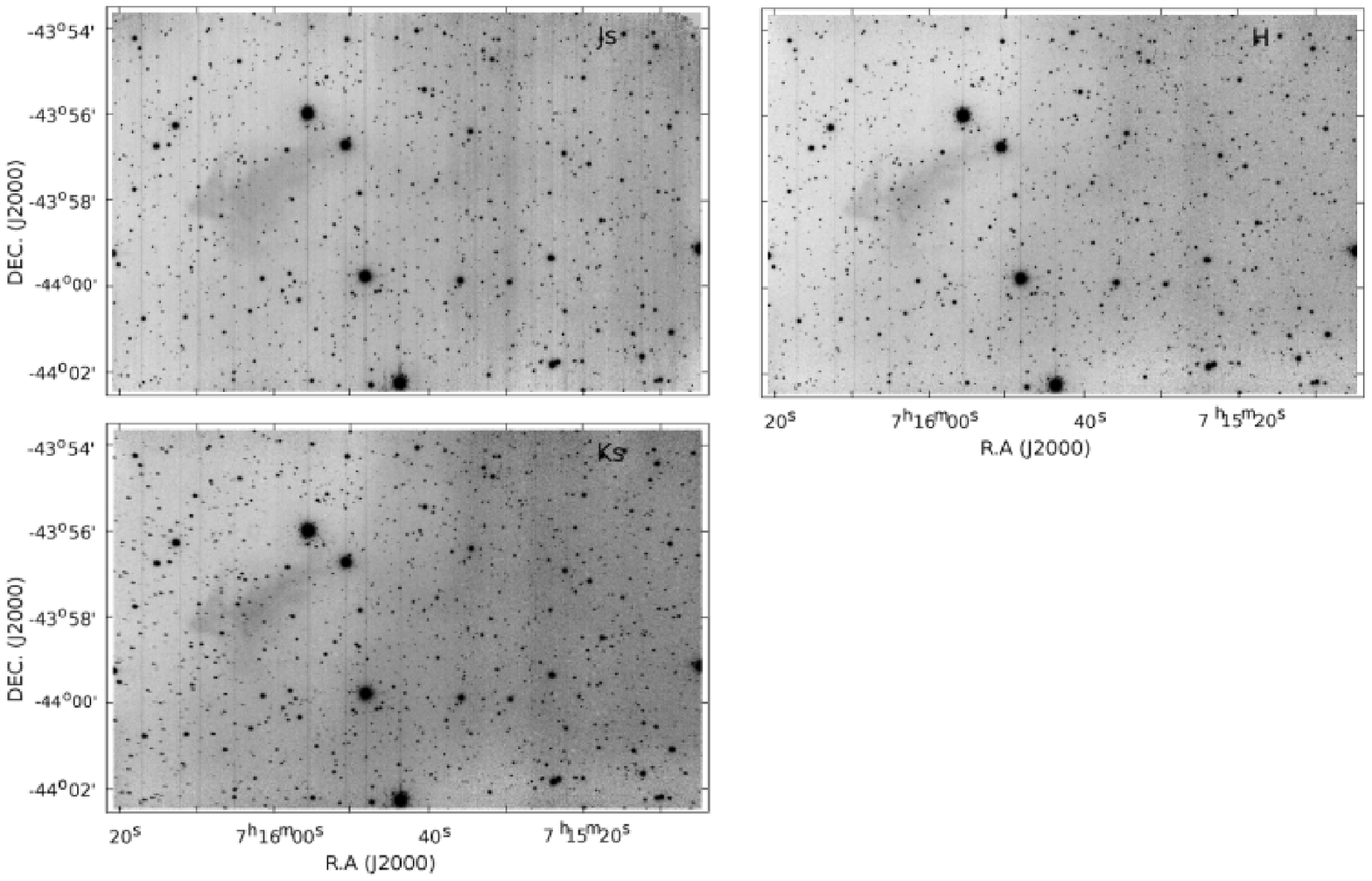}
   \caption{Sirius \J, \H, and \Ks\ images of CG~2.}
\label{fig:cg2sirius_all}
\end{figure*}

\begin{figure*}
\centering
\includegraphics [width=17cm] {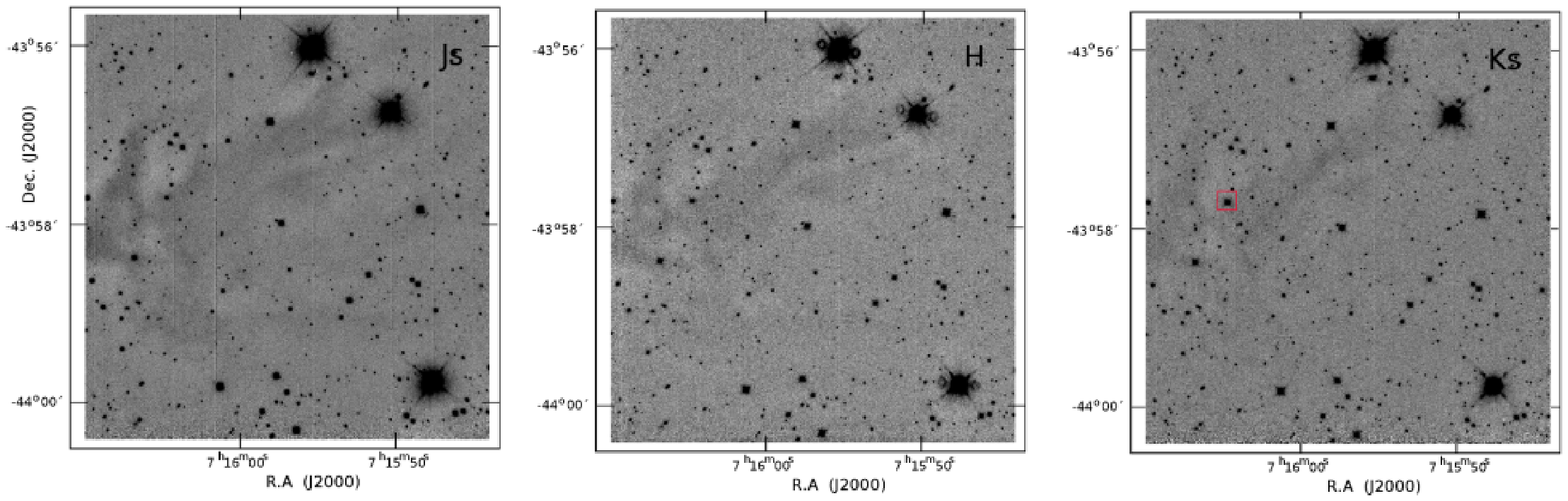}
   \caption{SOFI \Js, \H, and \Ks\ band images of CG~2.}
\label{fig:cg2_sofiall}
\end{figure*}

\begin{figure*}
\centering
\includegraphics [width=17cm] {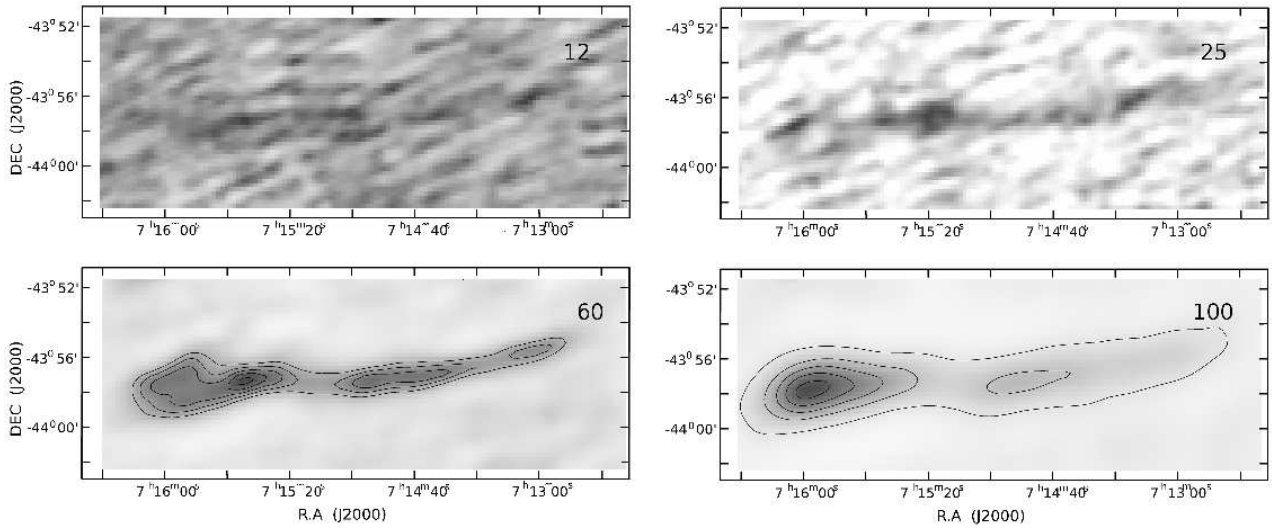}
   \caption{IRAS HIRES processed images of CG~2 in 12, 25, 60, and 100 $\mum$.}
\label{fig:cg2_irasall}
\end{figure*}

\begin{figure*}
\centering
\includegraphics [width=17cm] {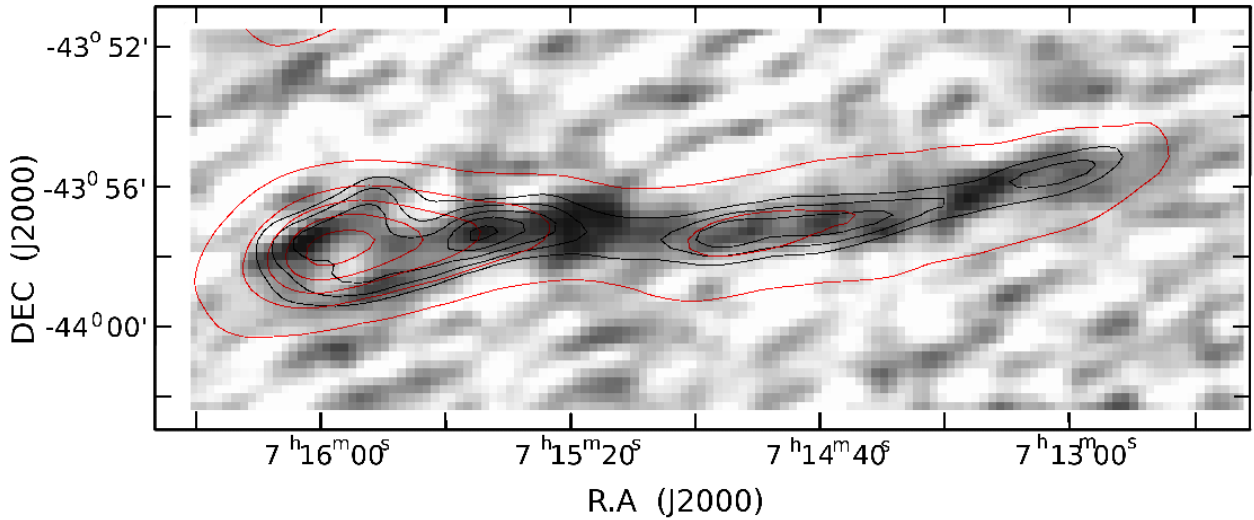}
   \caption{IRAS 25 $\mum$ HIRES processed image of CG~2. Overplotted are the IRAS HIRES processed 60 $\mum$ (black) and 100 $\mum$ (red) contours}.
\label{fig:cg2_w4_conwiseiras}
\end{figure*}

\begin{figure*}
\centering
\includegraphics [width=13cm] {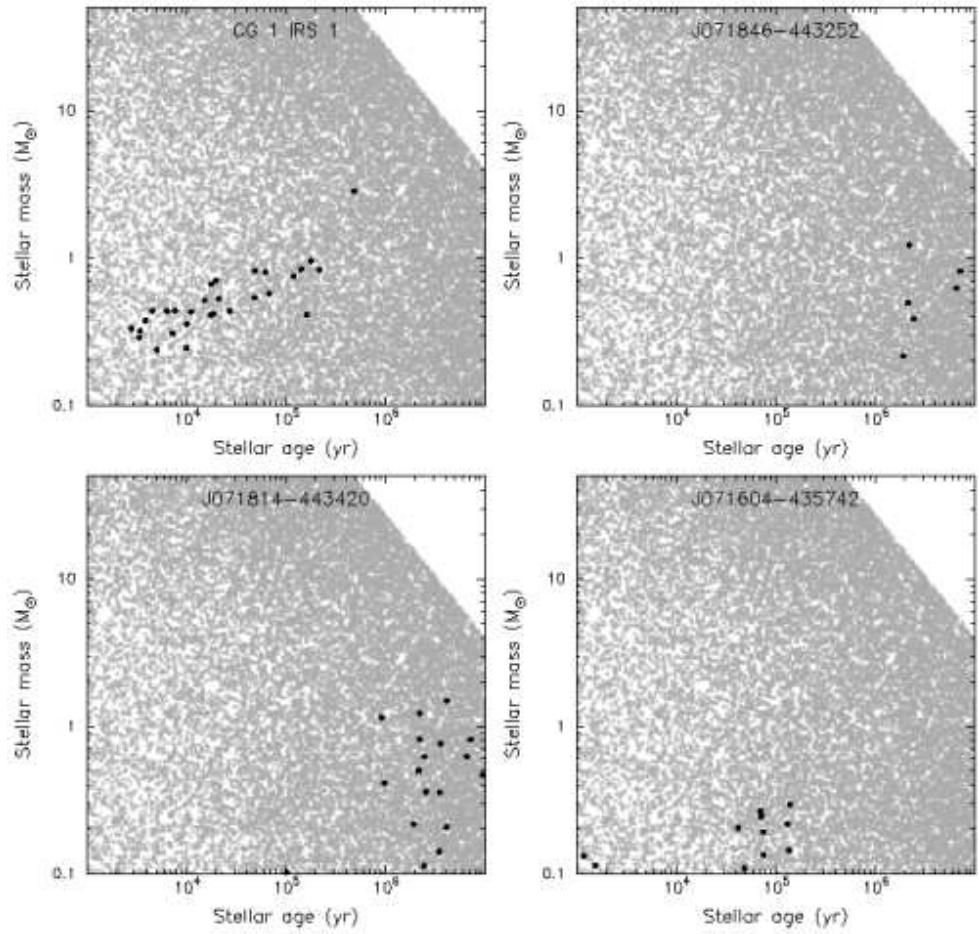}
   \caption{Age and mass in the SED fits. The filled grey circles show the distribution of models in the model grid, and the black filled circles shows the distribution of the best-fitting models i.e. with $\chi^{2}-\chi_{best}^{2} < 3$ per each data point.}
\label{fig:allsedfitsagemass}
\end{figure*}

\begin{figure*}
\centering
\includegraphics [width=17cm] {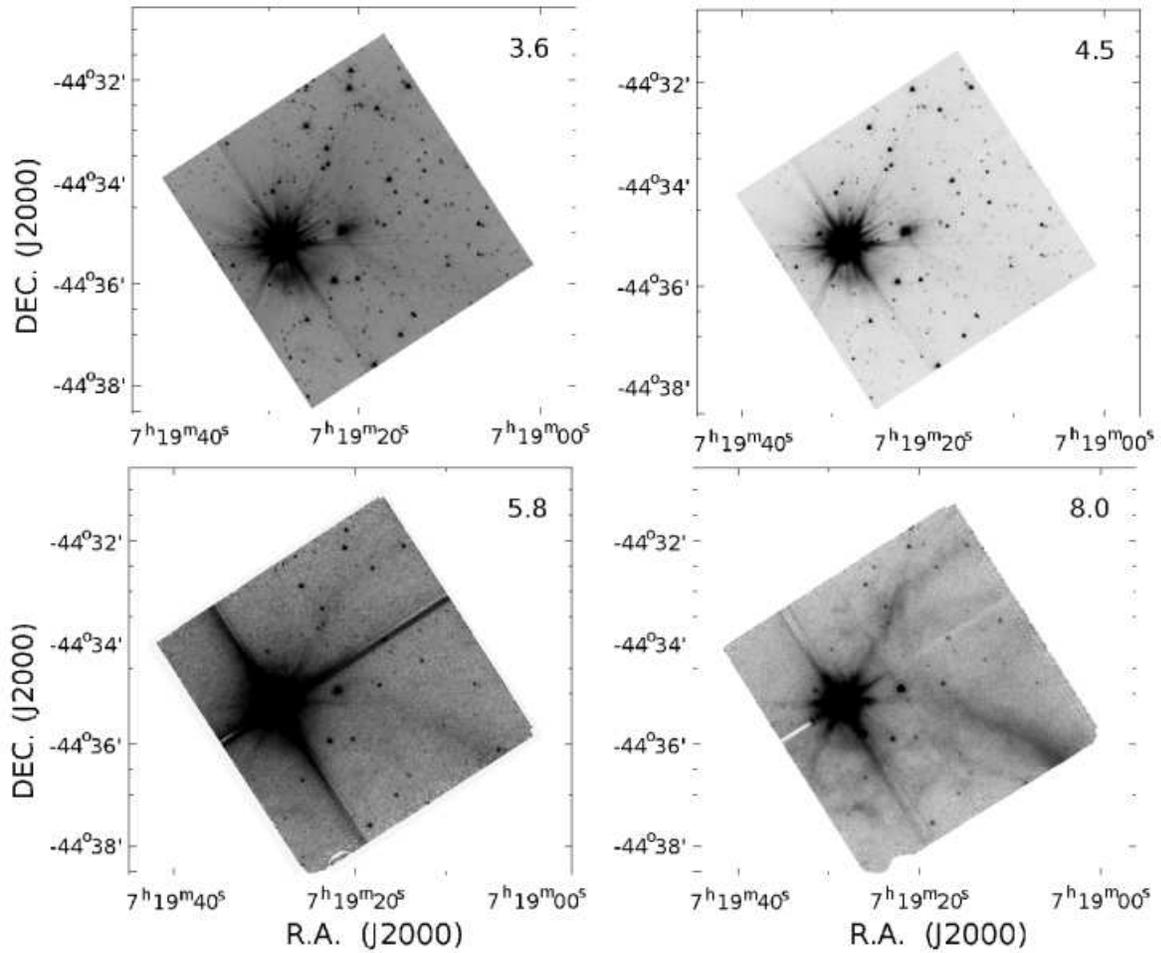}
   \caption{Spitzer IRAC 3.6, 4.5, 5.8, and 8.0 $\mum$ images of CG~1.}
\label{fig:cg1_iracall}
\end{figure*}

\begin{figure*}
\centering
\includegraphics [width=17cm] {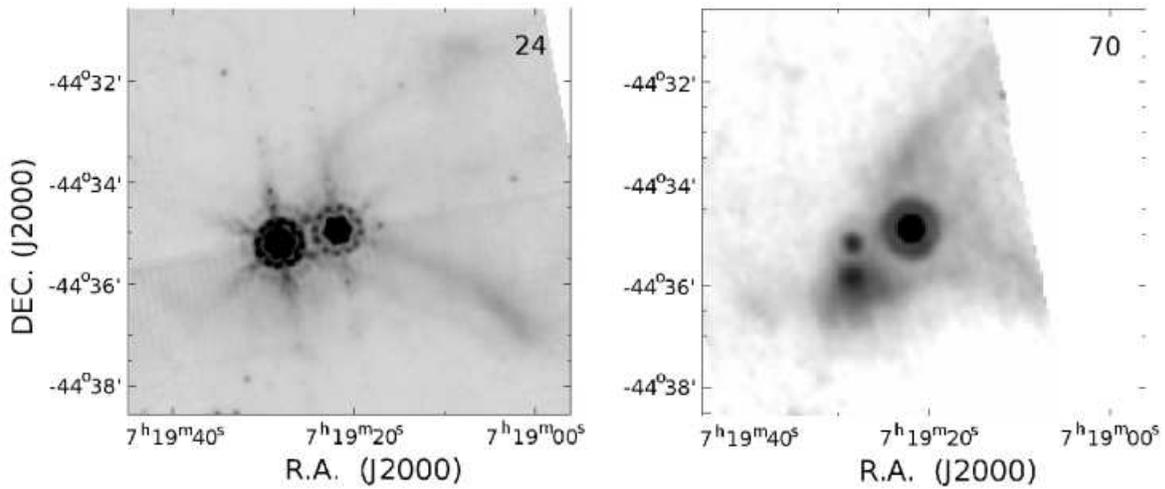}
   \caption{Spitzer MIPS 24 (left) and 70 $\mum$ (right) image of CG~1.}
\label{fig:cg1_spitzermips}
\end{figure*}

\begin{figure*}
\centering
\includegraphics [width=17cm] {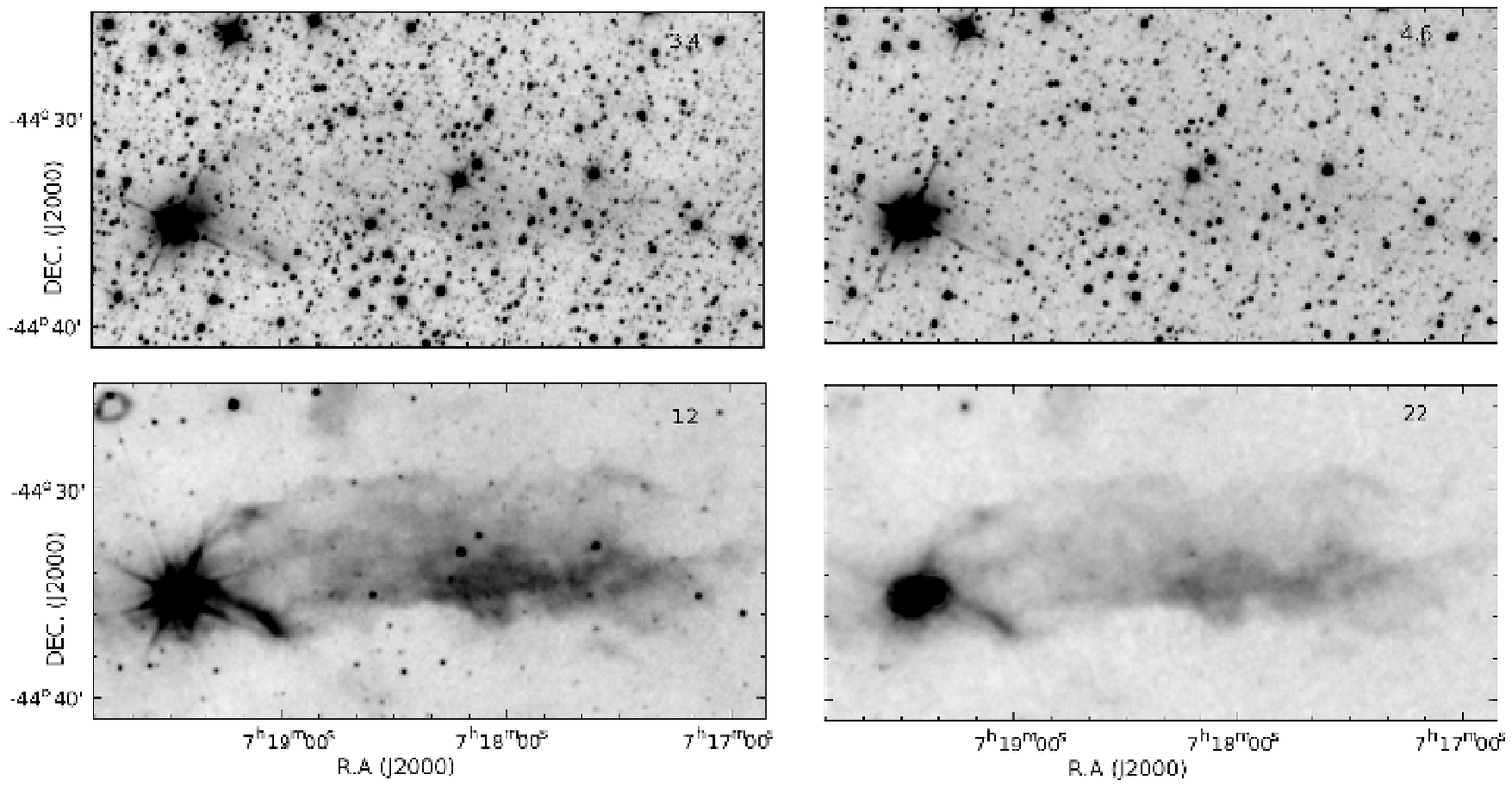}
   \caption{WISE 3.4, 4.6, 12, and 22 $\mum$ images of CG~1.}
\label{fig:cg1wise_all}
\end{figure*}

\begin{figure*}
\centering
\includegraphics [width=17cm] {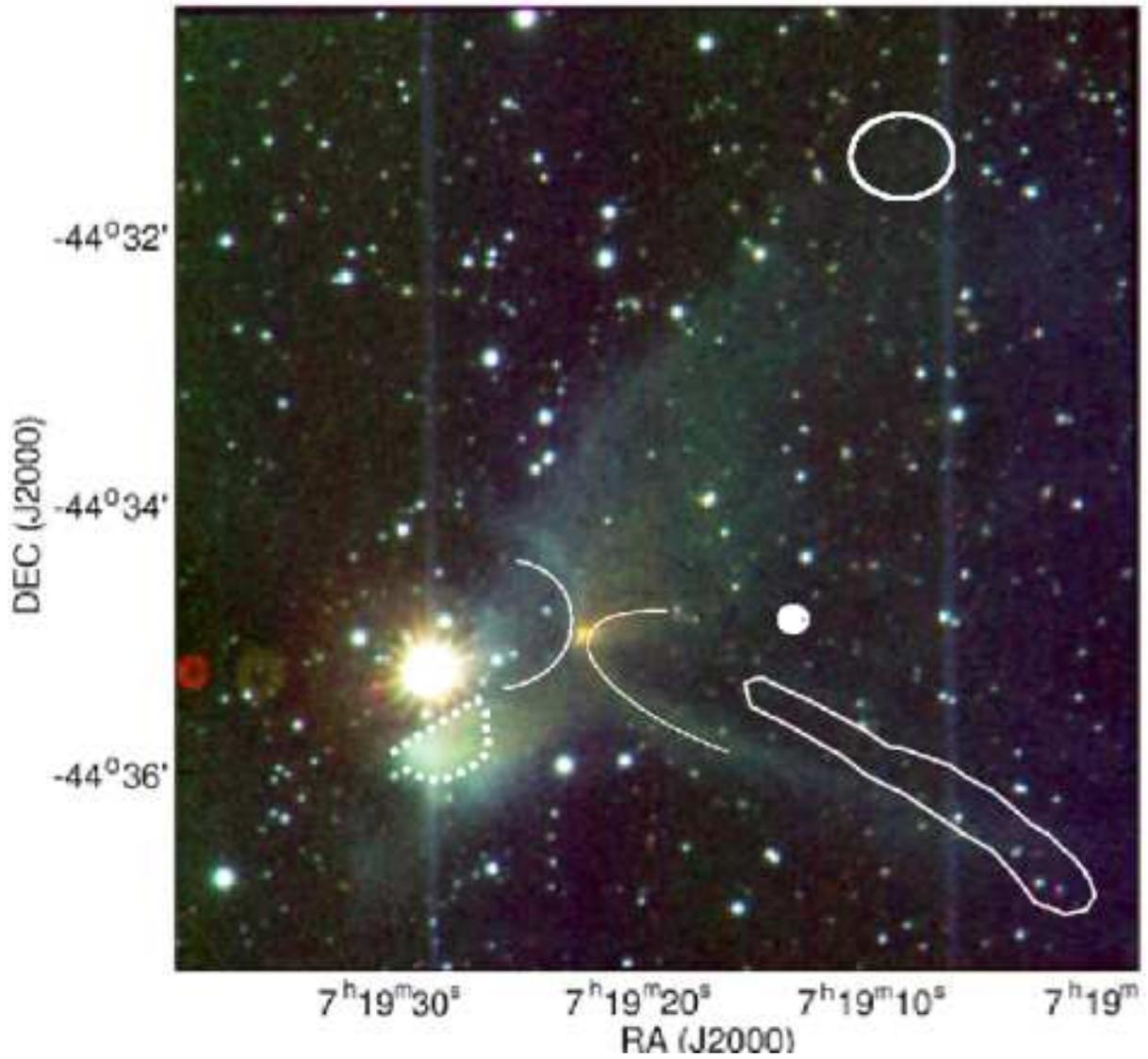}
   \caption{Observed structures in the CG~1 head over-plotted on the false-colour SIRIUS image. The thin, solid arcs starting from the YSO denote the possible outflow cones. The dashed semi-circle below the brightest objects NX Puppis marks the reflection nebula. The solid circle marks the location of the molecular hydrogen object MHO 1411. The ellipse in the top right-hand corner denotes the location of the patch of surface brightness, and the extended ridge (solid line) in the bottom right-hand corner marks the bright WISE\_SW streamer filament seen in the Spitzer images.}
\label{fig:cg1_structure}
\end{figure*}

\begin{figure*}
\centering
\includegraphics [width=17cm] {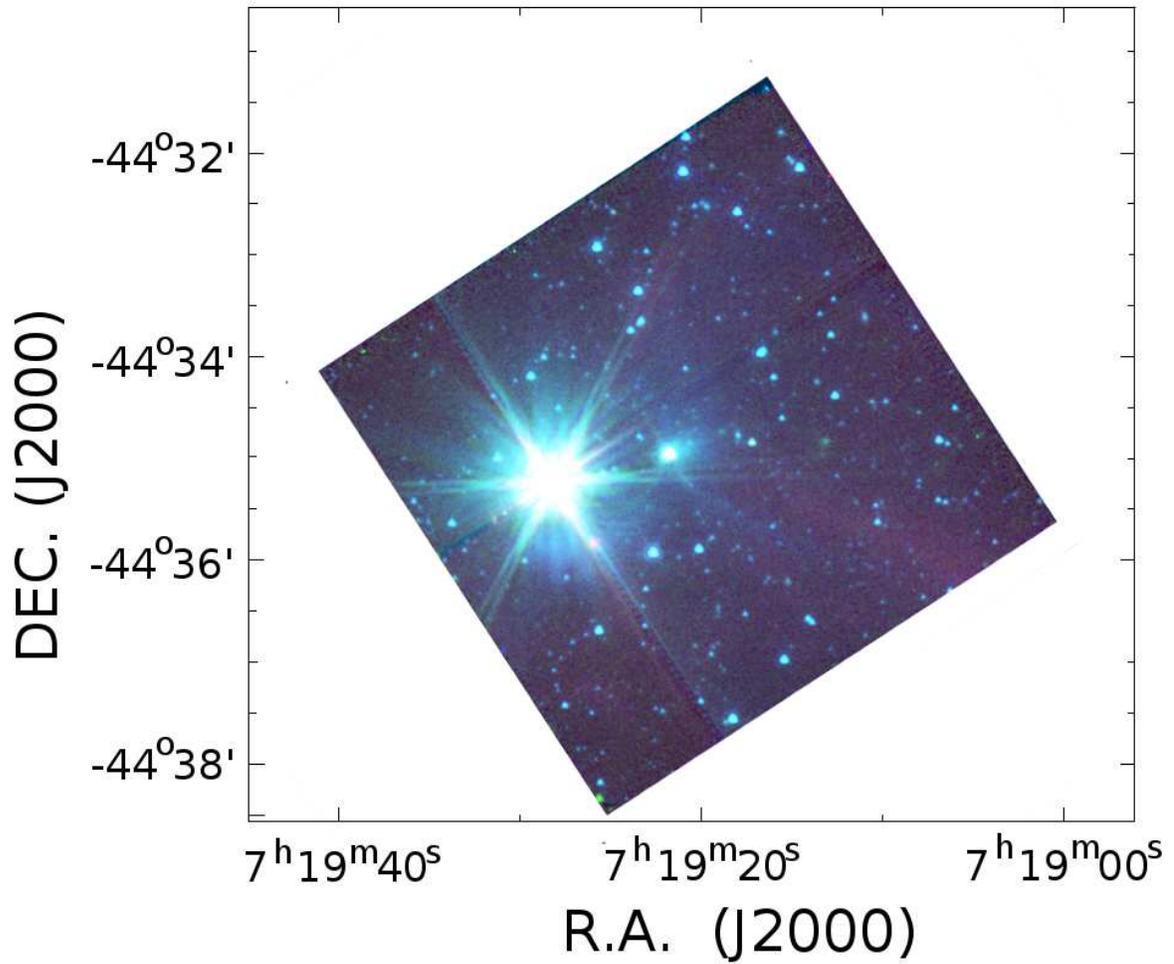}
   \caption{IRAC falsecolour image of CG~1. The 3.6 $\mum$ is coded in blue, 4.5 $\mum$ in green, and 8.0 $\mum$ in red.}
\label{fig:cg1_spitzer_color}
\end{figure*}

\begin{figure*}
\centering
\includegraphics [width=17cm] {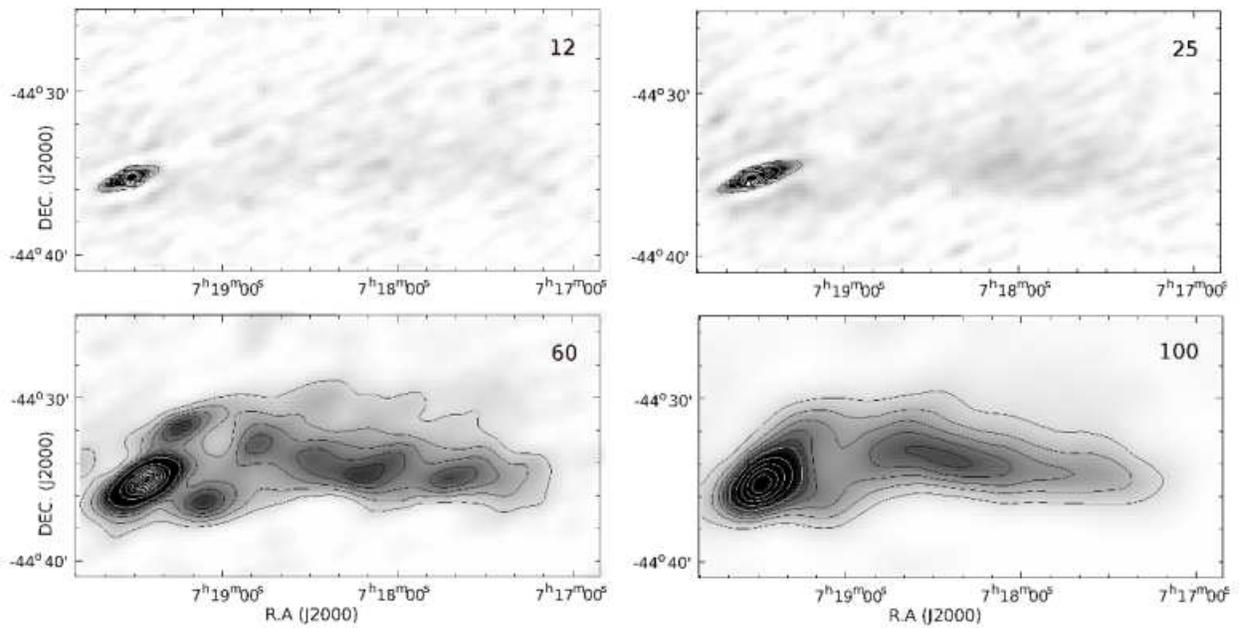}
   \caption{IRAS HIRES processed images of CG~1. Top left: 12 $\mum$. Top right: 25 $\mum$. Bottom left: 60 $\mum$. Bottom right: 100 $\mum$.}
\label{fig:cg1_irasall}
\end{figure*}

\begin{figure*}
\centering
\includegraphics [width=17cm] {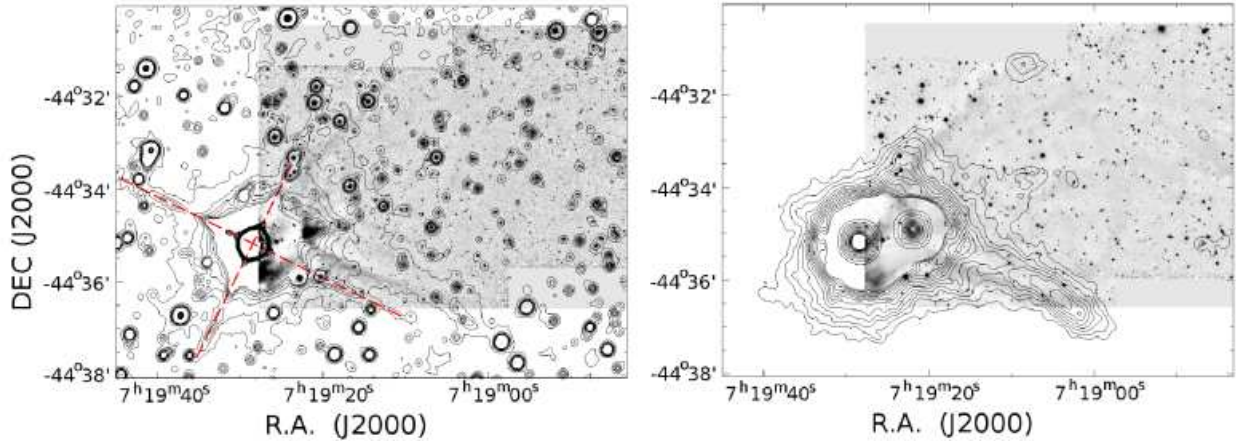}
   \caption{\Js\ image of CG~1. Left: The WISE 3.4 $\mum$ contours have been overplotted. The dashed lines indicate the strongest NX Pup diffraction spikes. Right: The WISE 22 $\mum$ contours have been overplotted. The contour levels are arbitrary.}
\label{fig:cg1_js_w1w4con}
\end{figure*}

\begin{figure*}
\centering
\includegraphics [width=17cm] {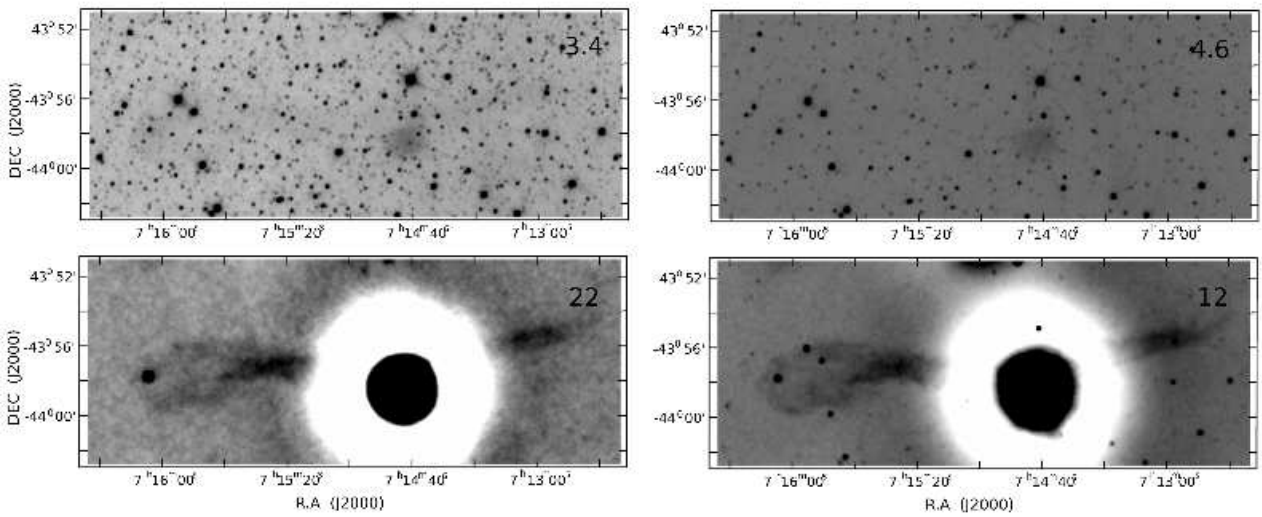}
   \caption{WISE CG~2 in 3.4, 4.6, 12, and 22 $\mum$ images. The strong artefact in the tail is caused by a nearby bright star.}
\label{fig:cg2_wiseall}
\end{figure*}

\end{appendix}

\end{document}